\documentclass{article}

\usepackage{PRIMEarxiv}

\usepackage[utf8]{inputenc} 
\usepackage[T1]{fontenc}    
\usepackage{hyperref}       
\usepackage{url}            
\usepackage{booktabs}       
\usepackage{amsfonts}       
\usepackage{nicefrac}       
\usepackage{microtype}      
\usepackage{lipsum}
\usepackage{fancyhdr}       
\usepackage{graphicx}       
\graphicspath{{media/}}     
\usepackage{natbib} 
\usepackage{rotating} 
\usepackage{threeparttable} 
\usepackage{amsmath} 
\usepackage{float} 
\usepackage{siunitx} 

\title{Portfolio Construction using Black-Litterman Model and Factors 
}

\author{
  Fanyu Zhao \\
  Website: \href{eightsmilefy.com}{fanyuzhao.com}\\
  E-mail \href{eightsmilefy.com}{eightsmilefy@gmail.com}
}

\begin{document}
\maketitle

\begin{abstract}
This paper presents a portfolio construction process, including mainly two parts, Factors Selection and Weight Allocations. For the factors selection part, We have chosen 20 factors by considering three aspects, the global market, different assets class, and stock idiosyncratic characteristics. Each factor is proxied by a corresponding ETF. Then, we would apply several weight allocation methods to those factors, including two fixed weight allocation methods, three optimisation methods, and a Black-Litterman model. In addition, we would also fit a Deep Learning model for generating views periodically and incorporating views with the prior to achieve dynamically updated weights by using the Black-Litterman model. In the end, the robustness checking shows how weights change with respect to time evolving and variance increasing. Results using shrinkage variance are provided to alleviate the impacts of representativeness of historical data, but there sadly has little impact. Overall, the model by using the Deep Learning plus Black-Litterman model results outperform the portfolio by other weight allocation schemes, even though further improvement and robustness checking should be performed. 
\end{abstract}


\section{Introduction} \label{Sec:Introduction}

In this paper, we would present a portfolio construction process by implementing mainly two steps. These are (1) Factor Selection, in which we would introduce the rationale of selecting factors building up our portfolio, and (2) Weight Allocation, in which we would assign weights to each factor. 

Appealingly, there are various methods of selecting factors and constructing a portfolio. Investors or fund managers may pick assets and assign weights through main three categories of ways, including (1) Heuristic Multi-factor Construction, (2) Optimised Multi-factor Construction, and (3) Risk-Based Multi-factor Construction \citep{nasdaq}. The first method, Heuristic Multi-factor Construction, is commonly used and easily to be used. Investors can simply choose assets they want and throw corresponding parameters and construct a linear model.

\begin{equation}
    \alpha_i = 0.2 F_{1,i} + 0.3 F_{2,i} + 0.5 F_{3,i} 
\end{equation} 

, where $F_{j,i}$ represents asset, with $i$ be the number in the sample, and $j$ be the number of chosen factor. Those $0.2$, $0.3$, and $0.5$ are just the arbitrary parameters, which are defined by investors, and those are measuring how $alpha$ is affected by factors $F_{j,i}$. However, there are potential drawbacks, such as high parameter bias, high factor selection bias, and lack of statistical evidence.

We would not implement the heuristic multi-factor construction method in our portfolio construction due to those above-mentioned drawbacks. However, we would learn from the Heuristic Model, selecting desired factors or assets to build our portfolio. My idea is to construct a factor-bearing portfolio, and we would use different ETFs or funds to proxy various factors. For example, the US growth equity could be captured by Vanguard Growth Index Fund. The rationale of choosing factors would be introduced in detail in the following sections.

Then, a combination of optimised multi-factor construction and risk-based multi-factor construction would be used to build up the portfolio. Specifically, we would use the minimum variance method, Markowitz mean-variance method, and reverse Sharpe ratio methods to construct the optimisation portfolio. Additionally, two fixed weight-allocation ways, namely equal weights and market cap weights, are also presented.  Also, the Black-Litterman Model has been used to dynamically adjust views on those select assets.

In this paper, we would go through a step-by-step work on portfolio construction. I would first introduce my logic of factor selection in Section \ref{sec:factor_selection}. We would apply several weight allocation methods to those factors, including two fixed weight allocation methods, three optimisation methods, and a Black-Litterman model. The methodology is introduced in Section \ref{sec:weights_allocation_methodology} and Section \ref{sec:BL_methodology}. In addition, we would also fit a Deep Learning model for generating views periodically and incorporating views prior to achieving dynamically updated weights by using the Black-Litterman model, as Section \ref{sec:LSTM}. Data, data source, and some statistics are shown in Section \ref{sec:data}. Results showing how weights are affected by different risk aversion scenarios and results showing the back-testing combining Deep Learning and the Black-Litterman model are presented in Section \ref{sec:results}. Later, the robustness checking shows how weights change with respect to time evolving and variance increasing. Results using shrinkage variance are provided in Section \ref{sec:robustness checking}. We at the end present also some discussions and improvements in Section \ref{sec:Discussion} and a conclusion section \ref{sec:Conclusion}.

\section{Methodology}

Before introducing the methodology and formula, let us introduce some the notations in our paper.

We aim to include 20 factors in our portfolio construction, each one is represented by a ETF or a fund. Plus, there is one benchmark, S\&P500 ETF and one risk-free return which is captured by the Treasury Bill rate. So, there are 22 assets being introduced, and each is a time series. We would use those 20 factors, 1 benchmark, and 1 risk-free rate to construct our portfolio.

\begin{itemize}
    \item  $w$: a vector of weights.
    \item $\mu$: a vector of the expected return for each factors
    \item $\Sigma$: a variance-covariance matrix for factors.
    \item $\mathbb{1}$: an one vector, with all elements being 1.
    \item  The superscript $\{T\}$ means the transport of a vector or a matrix. The superscript $\{-1\}$ means the inverse of a matrix.
    \item Matrix are mostly $N\times N$, where $N$ is the number of factors ($N=20$ in our case). However, views of the Black-Litterman model is a $K\times N$ matrix, where $K$ is the number of views, and $N$ is the number of factors, $N=20$ in our paper as there are 20 factors. Views are one-hot encoding.
\end{itemize}

\subsection{Factor Selection} \label{sec:factor_selection}

As I have introduced before, we are trying to construct a portfolio with different factors capturing different characteristics of the market. Three main rationales are considered to achieve the goal.

The first rationale is to consider the global market, including both the US and the emerging market. Our portfolio would include assets from the largest financial market, the US, and an emerging market where I would simply use the Chinese market, in order to diversify the risks in a global scope. Note that I would use China to represent the emerging market due to the fact that the Chinese financial market contains also a large volume of assets and personally, I  am more familiar with the market in China and can easily get access to the Chinese market. There are definitely disadvantages that we could not overcome, such as financial markets in China sometimes lack liquidity due to political implementations and China might not be that representative to all of other emerging markets. However, we would just follow our rationale. Also, to cope with the disadvantage, any other factor could be easily replaced in my portfolio construction framework.

The second rationale is to include different asset classes. Bonds, equities, and commodities are all considered in our portfolio because we are trying to capture most of the movements of the financial markets. Ideally, risks from different asset classes could be compensated, by combining and hedging. As the whole global market has been split into the US market and China market by our first rationale, we would now split further the market to be US bond, US equity, China bond, China Equity, and commodity. As the commodities are globally traded and globally priced, we cannot give a geographical separation to commodities. In addition, as suggested by the \textit{Reading Q\&A}, a VIX factor would be also included.

Thirdly, factors, such as growth factor, value factor, quality factor, small/large market cap factors, high-dividend factors, and factors that represent a specific industry are included for both the US and Chinese equities. There is much empirical financial evidence illustrating how those factors help capture the market movement. For instance, \citet{fama1993common} were pioneers who found the impacts of three well-known factors on financial assets, which are size effects, value effects, and the excess market. Then, \citet{fama2012size} stated that value, size, and momentum effects are largely existing in the global market. Inspired by findings from those scholars, we add those ETFs that mimic the above-mentioned factors into our consideration. In addition, both US technology and China technology ETFs are also added to my portfolio. We found the facts that the technology industry is one of the most iconic and high-growth industries for the US and Chinese markets, so we expect that the industry has certain representative characteristics that cannot be replaced by  other factors, and we also assume there is no perfect co-linearity between tech ETF and others. (Actually, certain factors are highly correlated but there are not perfect col-linearity as Figure \ref{fig:correlation} shows.)

Therefore, we pick the following 20 factors, as the \#1 to \#20 in Table \ref{tab:factor_list}. There are two more factors, \#0 and \#21 that are treated as the benchmark and the risk-free rate, respectively.

One common debate of factor selection is that there should be less correlation among those factors because factors could be better diversified. However, since we are considering different asset classes, we need a bit of active exposure to certain factors to a certain degree (see \textit{Reading Q\&A}).

\begin{sidewaystable}[htb]
\centering
\begin{threeparttable}
\caption{List of All Factors (22 Time Series, including 20 Factors 1 Benchmark and 1 Risk-free Rate}
\label{tab:factor_list}
\small 
\begin{tabular}{l l l l l l l}
\hline
\# & Type & yfinance Quote & Index Meaning & Variable Name & Factor \\
\midrule
0  & Equity US     & SPY            & SPDR S\&P 500 ETF Trust                       & sp500         & Benchmark \\
1  & Equity US     & VUG            & Vanguard Growth Index Fund                   & us\_growth     & US Growth \\
2  & Equity US     & VTV            & Vanguard Value Index Fund                    & us\_value      & US Value \\
3  & Equity US     & SPHQ           & Invesco S\&P 500 Quality ETF                  & us\_quality    & US Quality \\
4  & Equity US     & SPHD           & Invesco S\&P 500 High Dividend Low Volatility ETF & us\_dividend & US Dividend \\
5  & Equity US     & IYW            & iShares U.S. Technology ETF                   & us\_tech       & US High Tech \\
6  & Equity US     & VB             & Vanguard Small Cap Index Fund                 & us\_small      & US Small Cap \\
7  & Equity US     & MTUM           & iShares MSCI USA Momentum Factor ETF          & us\_momentum   & US Momentum \\
8  & Equity China  & MCHI           & iShares MSCI China ETF                        & china\_benchmark & China Benchmark Index \\
9  & Equity China  & ECNS           & iShares MSCI China Small-Cap ETF              & china\_growth  & China Growth \\
10 & Equity China  & FXI            & iShares China Large-Cap ETF                   & china\_value   & China Value \\
11 & Equity China  & CQQQ           & Invesco China Technology ETF                  & china\_tech    & China High Tech \\
12 & Equity China  & PGJ            & Invesco Golden Dragon China ETF               & china\_quality & China Quality \\
13 & Bond US       & SCHO           & Schwab Short-Term U.S. Treasury ETF           & us\_bond\_shortterm & Short-term US Bond \\
14 & Bond US       & TLT            & iShares 20+ Year Treasury Bond ETF            & us\_bond\_longterm & Long-term US Bond \\
15 & Bond China    & CBON           & VanEck China Bond ETF                         & china\_bond    & China Bond \\
16 & Commodity     & IEO            & iShares U.S. Oil \& Gas Exploration \& Production ETF & commodity\_oil & Oil and Gas \\
17 & Commodity     & IAU            & iShares Gold Trust                            & commodity\_gold & Gold \\
18 & Commodity     & DBA            & Invesco DB Agriculture Fund                   & commodity\_agriculture & Agriculture \\
19 & Real Estate   & VNQ            & Vanguard Real Estate Index Fund               & house\_us      & US Real Estate \\
20 & Volatility    & VIXY           & ProShares VIX Short-Term Futures ETF (VIXY)   & vix           & VIX Short-term \\
21 & Rate          & \textasciicircum IRX & 13 Week Treasury Bill                      & tbill         & 3-Month Treasury Bill Rate \\
\hline
\end{tabular}
\begin{tablenotes}[flushleft]
\small
\item Note: Table shows 22 Time Series that are used in this paper. There 20 Factors, each is represented by a ETFs, and  there are two more factors, Number 0 and Number 21. Those two are treated as the benchmark and the risk-free rate, respectively. Data would all retrieve from Yahoo!Finance.
\end{tablenotes}
\end{threeparttable}
\end{sidewaystable}

\subsection{Weight Allocation} \label{sec:weights_allocation_methodology}

\subsubsection{Fixed Weight Allocation Scheme}
Two different schemes of fixed weight allocation methods are used in this paper. The first method is the \textbf{equally-weighted method}. That means we assign equal weights to each factor. For example, each factor out of a total of 20 in total would be given weights $= 1/20 = 0.05 $ equally. The equal-weighted portfolio is easy to be constructed, but is hard to tackle the problem when investors have their own desire on certain factors, or when investors have different risk tolerance levels.

The other method is the \textbf{market capitalisation weighted method}, in which we assign weights by using the percentage of market cap for each factor. As all factors have their corresponding ETFs, the market capitalisation data could be retrieved from data providers such as Yahoo! Finance. However, there are some problems we are encountering while using the percentage of market cap to allocate weights to each factor. The first problem is ETFs, as the proxies of those factors are produced by different fund providers, investors have different desires for those ETF providers. Meanwhile, there are many ETF providers that provide similar ETFs that proxy the same factor, such as small-cap ETFs. In this case, using ETFs market cap to calculate the market capitalisation weighted portfolio would give a distorted result. Another problem is that using the market cap to weight factors is based on the assumption that the market is already in equilibrium. Although the "efficient market" assumption is generally used, there is also empirical evidence against the assumption.

The third problem is that there are some technical issues we are undertaking while doing this assignment. The main database we are using, Yahoo!Finance, is under maintenance, so that market cap data are unavailable at that time. We would program an algorithm retrieving historical time series data of market caps for each factor, once Yahoo!Finance is fixed. Also, in real practice, I definitely would switch to other data providers or exchanges.

\subsubsection{Weights from Optimisation}
Four optimisation methods are employed to establish the portfolio in our paper, namely (1) the global minimum variance portfolio, (2) the maximum Sharpe ratio portfolio, (3) the Markowitz mean-variance portfolio, and finally (4) the reverse-beta portfolio. They all employ certain optimisation functions given by different constraints to calculate weights, and we would introduce them in detail as follows.

\subsubsection{Global Minimum Variance Portfolio}
The global minimum variance portfolio has the objective function of minimising the portfolio variance. As Equation \ref{eq:gmv} shows, the portfolio variance is calculated as $w^T \Sigma w$.

\begin{equation}
    w_{GMV} =arg\min_w w^T \Sigma w \label{eq:gmv}
\end{equation}

$$s.t.\quad  \mathbb{1}^T w = 1$$

Solve it by taking the first order condition (F.O.C.) to the Lagrangian equation.

\begin{equation}
    w_{GMV} = \frac{\Sigma^{-1} \mathbb{1} }{\mathbb{1}^T \Sigma^{-1}\mathbb{1}} \label{eq:GMV_sol}
\end{equation} 

, where $w_{GMV}$ is the Global Minimum Variance Portfolio Weights.

We do apply constraint on weights to restrict them sum to be one, and in a range of $[0,1]$.

\subsubsection{Maximum Sharpe Ratio Portfolio}
\begin{equation}
    w_s = arg\max_w \frac{w^T (\mu - rf)}{ \sqrt{w^T \Sigma w} } \label{eq:maxSharpe}
\end{equation}  

$$s.t. \mathbb{1}^T w = 1$$

By solving Equation \ref{eq:maxSharpe}, we would obtain the solution, as the following.
\begin{equation}
    w_s = \frac{\Sigma^{-1}\mu}{\mathbb{1}^T \Sigma^{-1} \mu} \label{eq:maxSharpe_sol}
\end{equation}

We do apply constraint on weights to restrict them sum to be one, and in a range of $[0,1]$.

\subsubsection{Markowitz Mean-Variance Portfolio}

\begin{equation}
    w_m = arg\max_w \big\{w^T \mu - \lambda w^T \Sigma w \big\} \label{eq:Markowitz}
\end{equation} 

$$s.t.\quad  \mathbb{1}^T w = 1$$

The Lagrangian would thus be, 

\begin{equation}
    \mathcal{L} = w^T \mu - \lambda w^T \Sigma w + v (1-w^T\mu)
\end{equation}

Solve by taking the F.O.C. on the Lagrangian equation, again.

$$
    \frac{\partial \mathcal{L}}{\partial w} = \mu - 2\lambda \Sigma w - v\mathbb{1} \equiv 0
$$

$$
    w = \frac{1}{2\lambda} \Sigma^{-1} \big[ \mu - v\mathbb{1} \big]
$$

, by substituting that $w$ into the constraint $\mathbb{1}^T w = 1$

$$
    v = \frac{\mathbb{1}^T \Sigma^{-1} \mu -2 \lambda }{\mathbb{1}^T \Sigma^{-1}\mathbb{1}}
$$

Substitute into $w$ 
 
\begin{equation}
\begin{split} 
w &= \frac{1}{2\lambda} \Sigma^{-1} \big[ \mu - v \mathbb{1}  \big] \\
  &= \frac{1}{2\lambda} \Sigma^{-1} \big[ \mu - \frac{\mathbb{1}^T \Sigma^{-1} \mu -2 \lambda }{\mathbb{1}^T \Sigma^{-1}\mathbb{1}} \mathbb{1}  \big]
\end{split}
\end{equation}

We do apply constraint on weights to restrict them sum to be one, and in a range of $[0,1]$.

\subsubsection{Implied Beta Weighted Portfolio}
The Implied Beta Method is proposed by \citet{kahn1999active}, and then applied by \citet{idzorek2007step}. \citet{idzorek2007step} employ the \textit{implied beta} method to calculate the implied betas. Then, Idzorek apply the implied betas, instead of regression base betas, to obtain the implied equilibrium return of the market. 

In our paper, as the time-series data of market capitalisation are not available, we made a bit modification of the implied betas method, in order to provide an alternative way to find weights for each factor. We firstly get the time-series data of each factor, and calculate the regression base beta by using the benchmark, S\&P500. Then, using the vector of beta to reversely obtain the vector of weights for each factor. Here below is the derivation.

Firstly, we apply the CAPM way to calculate the $\beta_i$, $i$ is the factor. 
$\beta$ in the CAPM is expressed as, 

$$ R_i - r_f = \beta_i \times (R_m - r_f) $$

$$ \beta_i = \frac{ Covariance(R_i, R_m) }{ Variance(R_m) } $$

Secondly, we reverse the formula provided by \citet{idzorek2007step} and \citet{kahn1999active}
to get the "implied beta weights",

$$\beta = \frac{ \Sigma w_{\beta} }{w_{\beta}^T \Sigma w_{\beta}} = \frac{\Sigma w_{\beta}}{\sigma^2} $$

\begin{equation}
    w_{\beta} = \sigma^2  \Sigma^{-1} \beta  \label{eq:impliedBeta}
\end{equation}

, where

\begin{itemize}
    \item  $\beta$: is the vector of implied betas; $\beta_i$ is the scalar of beta for a certain factor;
    \item  $\Sigma_{rf}$: is the covariance matrix of excess returns, excess to the risk-free rate;
    \item  $w_{\beta}$: is the implied beta weights we have just illustrated;
    \item $\sigma^2 = w_{\beta}^T \Sigma w_{\beta} = \frac{1}{\beta^T \Sigma^{-1} \beta}$ is the variance of the market (or benchmark) excess returns. 
\end{itemize}

Finally, we can get the $w_{\beta}$,

We do not apply constraints here.

\subsection{Black-Litterman Model} \label{sec:BL_methodology}
By using the Black-Litterman, we could incorporate the view with the prior, and  then get the posterior. Let's now formulate the Black-Litterman Model.

\subsubsection{The Prior}

We first define the prior distribution. We assume the excess return as a random variable follows a normal distribution, with mean $\mu$ and variance $\Sigma$ as the following,

$$ r \sim N(\mu,\Sigma) $$

We denote $\mu$ here as an unknown "actual" mean return in this section, and $\mu$ is our \textbf{prior}.

$$ \mu \sim N\Bigg(\mathbb{E}(\mu),\Sigma_{\pi}\Bigg) $$

, where $\pi = \mathbb{E}(\mu)$, and $\mu = \pi +\epsilon$. $\epsilon \sim N(0,\Sigma_{\pi})$. So the distribution of the prior could be rewritten as,

$$ \mu \sim N\bigg(\pi,\Sigma_{\pi}\bigg) $$

\begin{itemize}
    \item $\pi$ is our estimate of the mean, or in other words, it is the expected excess equilibrium return.
    \item $\Sigma_{\pi}$ is the variance of the estimate.
    \item The $\pi$ here is what we can estimate from the market.
\end{itemize}

\paragraph{The Equilibrium Return}

As inspired by \citet{walters2014black}, we would calculate the risk-aversion first.

\paragraph{The Risk Aversion in the Mean-Variance Optimisation }

Let's consider a situation investing in either the ready-picked risky portfolio or a risk-free asset. 

We denote,
\begin{itemize}
    \item  $\mu_B$ as the benchmark excess return, excess to the risk-free rate.
    \item  $\sigma_B$ as the benchmark risk.
    \item  $w_B$ be the weights investing risky assets.
    \item  $1-w_B$ the rest would be investing in risk-free assets.
    \item  $\lambda_{mkt}$ denotes the risk aversion of the market, with is the trade-off between risky and risk-free assets.
    \item  P.S. the excess return and excess risk for the risk-free asset are both zero.
\end{itemize}

$$ w_B = arg\max_w \bigg\{ w_B \cdot \mu_B - \lambda_{mkt} (w_B \sigma_B)^2 \bigg\} $$

By F.O.C. w.r.t. $w_B$, we will get,

$$ \mu_B - 2\lambda_{mkt} w_B \sigma^2_B = 0 $$

As we suppose the benchmark portfolio is the optimal portfolio, we move all our weights to the risky asset, let $w_B \equiv 1$.

\begin{equation}
    \lambda_{mkt} = \frac{\mu_B}{2\sigma^2_B} \label{eq:lambda_mkt}
\end{equation}

Therefore, we would get $ \lambda_{mkt} = \frac{\mu_B}{2\sigma^2_B} $ as our Risk Aversion.

We would proxy the market benchmark by the S\&P500 index, and proxy the risk-free return by the 3-month Treasury Bond rate, which is one of the most risk-free assets in the market.

\paragraph{Solve the Equilibrium Return by Reverse Optimisation}

Consider Markowitz's Optimisation problem. Since we are dealing with a market portfolio that has only positive weights and those weights are summing to be 1, we would not implement constraint on that. The objective function is as the following.

$$ arg\max_w \big\{ w^T \pi - \lambda_{mkt} w^T \Sigma w \big\} $$

by F.O.C.

$$\frac{\partial .}{\partial w} = \pi - 2\lambda_{mkt} \Sigma w \equiv 0$$

$$w = \frac{1}{2\lambda_{mkt}} \Sigma^{-1} \pi$$

, or the reverse,

\begin{equation}
    \pi = 2\lambda_{mkt} \Sigma w  \label{eq:weight2pi}
\end{equation} 

$$$$
, where 
\begin{itemize}
    \item  $\pi$ is a vector of the equilibrium excess return;
    \item  $\lambda_{mkt}$ is the empirical risk aversion derived from the benchmark (in our case is S\&P500)
\end{itemize}

Equation \ref{eq:weight2pi} provides us with a way to calculate "returns" through weights (Inputting a vector of weights and outputting a vector of returns). For example, if we assume the market cap weights could represent the equilibrium situation of the market, then passing the market cap weights into Equation \ref{eq:weight2pi} would give us the implied equilibrium return. 

\subsubsection{Posterior -  The Black Litterman Formula}

As noted by \citet{da2009black}, we are now focusing on the excess return over the benchmark, $\pi$ instead of the risk-free return, $r_f$. Let's now implement the active assets management that maximises the expected excess return, $\mu_{BL}$.

The Black Litterman model could be expressed as,

$$ \mathbb{E}[R] = \mu_{BL} = \bigg[ (\tau \Sigma)^{-1} + P' \Omega^{-1} P \bigg]^{-1} \bigg[ (\tau \Sigma)^{-1}\pi + P' \Omega^{-1} Q \bigg] $$

, where
\begin{itemize}
    \item  $\mathbb{E}[R] = \mu_{BL}$ is an $N \times 1$ vector of Black-Litterman expected excess returns, and it is also the Posterior.
    \item  $\pi$ is the equilibrium excess return;
    \item  $\tau$ is the scaling factor representing the uncertainty in the market equilibrium;
    \item  $\Sigma$ is an $N \times N$ variance-covariance matrix of assets' excess returns, 
    \item  $P$ is a $K \times N$ matrix that has $K$ views on $N$ assets; 
    \item  $\Omega$ is the matrix that represents the conﬁdence in each view, and 
    \item  $Q$ is a $K \times 1$ vector of expected returns of those K views. 
\end{itemize}

A view portfolio may include one or more assets through nonzero elements in the corresponding elements in the P matrix.

\subsubsection{Views}
A core concept and an important application of the Black-Litterman Model is that the model can update Views. In other words, Views are incorporated with the Prior, and then we can deduce the Posterior.

Views reflect investors' information and belief about the further movement of the market \citep{maggiar2009active}. Views can be categorised as either Absolute or Relative views, and either Asset Specific or Global views.

\subsubsection{Types of the Views}

\paragraph{Absolute v.s. Relative views}
    An absolute view states the absolute level of expected excess return for an asset. For example, the expected excess return of asset A is 2\%.
    
    In contrast, the relative view states the expected excess return outperforms or under-performs the other. In the case of relative views, all elements in a row are summed to 0. For instance, the expected excess return of asset A will outperform asset B by1 1\%. 

\paragraph{Asset Specific v.s. Global views}

    The Asset Specific view states the view for only one asset. However, the global view expresses views on a set of assets. For example, a portfolio consisting of asset A and asset B would outperform the portfolio consisting of asset C and asset D.

We denote $P$ as a $K \times N$ matrix, where $K$ is the number of views and $N$ is the number of assets. Each raw represents a view. Based on the example of \citet{RePEc:pal:assmgt:v:1:y:2000:i:2:d:10.1057_palgrave.jam.2240011}, we raise an example in our case,

\begin{equation}
P = \bigg( \begin{array}{*{20}{c}}
   0 & 0 & 0 & 0 & 0 & 0 & 1 & 0 & 0 & 0 & 0 & 0 & 0 & 0 & 0 & 0 & 0 & 0 & 0 & 0 \\
   1 & -1 & 0 & 0 & 0 & 0 & 0 & 0 & 0 & 0 & 0 & 0 & 0 & 0 & 0 & 0 & 0 & 0 & 0 & 0 \\
   0 & 0 & 0 & 0 & 0 & 0 & 0 & \frac{1}{5} & \frac{1}{5} & \frac{1}{5} & \frac{1}{5} & \frac{1}{5} & 0 & 1 & 0 & 0 & 0 & 0 & 0 & 0 \\
\end{array} \bigg)
\label{eq:views}
\end{equation}

Also, $q$ as a $K \times 1$ vector represents the view results.

$$q = \begin{pmatrix}  0.01 \\ 0.01 \\ 0.02  \end{pmatrix} $$

There are three views in our example. 
\begin{itemize}
    \item The first row of matrix $P$ is an example of the absolute view and also a specific view. Only the `us\_momentum` would be expected to generate a 1\% excess return.
    \item The second row of matrix $P$ represents the relative view and is specific to `us\_growth` and `us\_value`. The view states the return of holding `US Growth Stocks` would outperform using the `US Value Stocks` index by 1\%. 
    \item The third row of matrix $P$ is a relative and global view, stating all US stocks would together outperform Chinese Stocks, by overall 2\%.
\end{itemize}

\subsubsection{Uncertainty of the Views}

We use $\Omega$ to denote the uncertainty matrix of views. By assuming views are independent to each other, the variance-covariance and correlation between views are all zero, so that $\Omega$ is a diagonal matrix. The off-diagonal elements of $\Omega$ are all zero because the model assumes that views are independent to others.

Although the assumption that views are independent might not be realistic because sometimes investors' views have logistic induction from one view to the other. However, we would not discuss correlated views in this paper. Correlated views are easy to be implemented by changing values in the off-diagonal matrix, $\Omega$.

There are two types of $\Omega$ matrix used by scholars and practitioners. The first type of matrix is the following. Diagonal elements $\omega_i^2$ imply the uncertainty level of the $i$'s view.

$$\Omega_{K\times K} = \begin{pmatrix}  \omega^2_1 & 0 & \cdots & 0 \\ 0 & \omega_2^2 &\cdots &  0 \\ 0 & 0 & \ddots & 0\\ \vdots & 0 & 0 & \omega^2_k  \end{pmatrix} $$

The second type of view is the following equation. In this case, we assume a linear relationship between the variance for all factors ($\Sigma$) and the variance for the views ($\Omega$).

$$\Omega = P \tau \Sigma P^T$$

\subsection{Deep Learning Algorithm - LSTM} \label{sec:LSTM}
The main purpose of the Black-Litterman model is to incorporate views with the prior, and then generate the posterior. We would program a deep learning algorithm to periodically update views. 

In our model, data would be rolled and split into several time spans, because we set our task to generate views periodically. We aim to create a classifier that uses previous data to predict a view instructing investors which factor could generate excess returns over the following period. In that case, we would train a new neural network for each time span, in order to make our views periodically updated.

\begin{figure}[htb]
  \centering
  \includegraphics[width=1.0\textwidth]{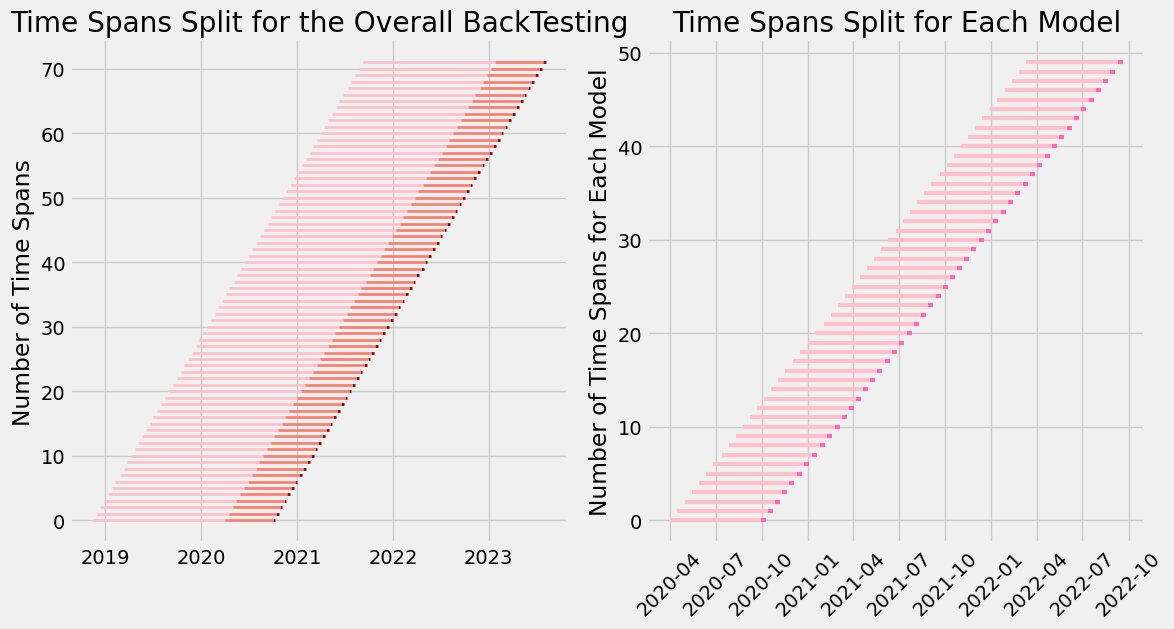}
  \caption{Time Spans Split}
  \label{fig:TimeSpanSplit}
\end{figure}

Figure \ref{fig:TimeSpanSplit} shows how the time spans are split. 

\paragraph{Here is what the left-subplot shows.}
The \textbf{left sub-plot} shows how each model is trained and how the loop is performed. Each horizontal line is a single round in the loop. We split each horizontal line into three parts.

\begin{itemize}
    \item The \textit{light-pink} part is used to train an LSTM model. Every single \textit{light-pink} bar includes $126\times 4 = 504$ days. Those parts of the data are employed to train a LSTM model in that round of the loop. 
    \item The \textit{darker} part from each bar is a sub-sample used to calculate weights under different weight allocation schemes. The \textit{darker} part would also be an input into the pre-trained model using \textit{light-pink} part.
    \item The \textit{darkest} part contains 10 days. That part is the output of view, after inputting the \textit{darker} part into the pre-model. Also, that part is the duration we invest our re-balanced weights.
\end{itemize}

\paragraph{Here is what the right-subplot shows.}
The \textbf{right-subplot} shows how a single \textit{light-pink} part in the \textbf{left-subplot} could be sub-decomposed. a \textit{light-pink} bar in the \textbf{left-subplot} is decomposed into several sub-samples, rolling each 10 days. Then those decomposed data become the inputs of an LSTM model. In the \textbf{right-subplot}, the \textit{feature} is pink part includes $126$ days of data, and the \textit{target} is calculated from the \textit{darker-pink} part. Note that we do not perform a train-validation split for each pink bar. We do not care too much about how reliable the LSTM model is, because it is too subtle and is just a tool to generate views. That is for the paper writing purpose only, a more consolidated model would be replaced in real practice.

 Each time span has $126$ days, as the market regimes could switch so we sometimes do not want to incorporate all the historical data. We would like to choose one factor that can generate the highest cumulative return over the next 10 days. The time span would be rolled each 10 days as well. Thus, the LSTM model is trained by using $N\times 126\times 20$ (N sub-samples, 126 days and 20 factors) as the input \textbf{features}, and use the argmax label of the next following 10-day cumulative return of each factor as the output \textbf{target}. We re-balance all weights every 10 days. 

 The view would be transformed into one-hot encoding, which is exactly the same as the format of $P$ matrix, while only one view is generated for each round of the loop. We manually set the $Q = 0.01$ as a hyper-parameter, and $\tau = 1/252$ denoting the one-day level certainty.

We perform the loop in this way can ensure the train-set of each LSTM never overlap with its predicted view. 

Hyper-Parameters of our model are:
\begin{itemize}
    \item windows = $10$ days
    \item sequence\_length = $126$ days (, which is the size to a sub-sample)
    \item the number of data input to train an LSTM model = $4\times 126 = 504$ days. So the shape of data input would be $504/10 \times 126 \times 20 = 50 \times 126 \times 20$.
    \item $Q=0.01$
\end{itemize}

\section{Data} \label{sec:data}
All the data we use in this paper are extracted from Yahoo!Finance. As we have mentioned in the weight allocation part, market capitalisation data is currently not available, so we manually scrap those data from Yahoo!Finance. Algorithm could be easily performed once Yahoo!Finance is fixed, and data from any other data providers could be easily applied to my algorithm once transforming data into "list", "pandas.Dataframe", "numpy.arrary" and "pandas.Series".

The whole list of factors is shown as Table \ref{tab:factor_list}. The sample space includes panel data of 22 time series (20 factors, 1 benchmark, and 1 risk-free rate) from the 1st of April 2020 to the 1st of August 2023, daily. The summary statistics are shown in Table \ref{tab:sum_statistics} below. Figure \ref{fig:cum_return} shows the cumulative returns for each factor.

\begin{figure}[htb]
  \centering
  \includegraphics[width=0.7\textwidth]{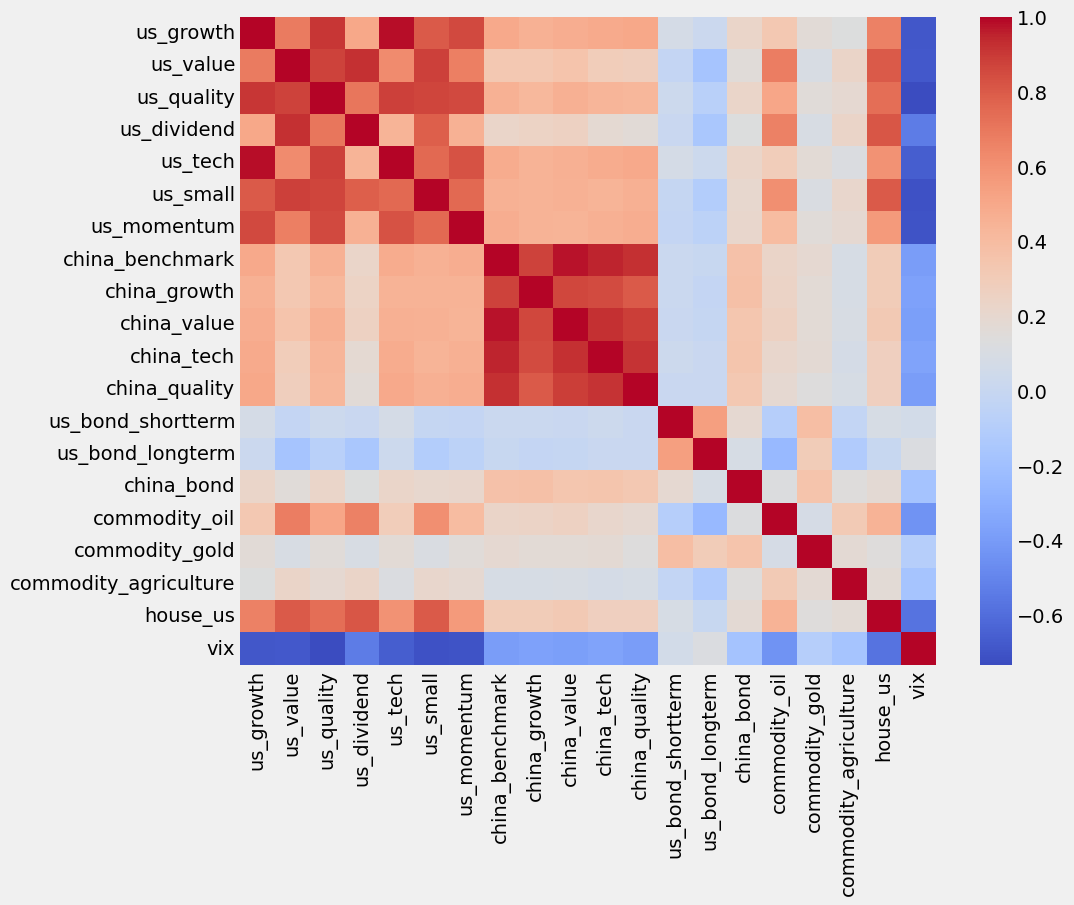}
  \caption{Correlation}
  \label{fig:correlation}
\end{figure}

Figure \ref{fig:correlation} shows the correlation between each factor. The warmer the colour, the more correlated that pair of factors is. The cooler the colour, the less correlated the pair is. From Figure \ref{fig:correlation}, we could find the US Stocks (, including us\_growth, us\_value, us\_quality, us\_dividend, us\_tech) are highly correlated with each other. Chinese Stocks are also highly correlated. Bonds have a negative correlation with Equity from each market. Oil is positively correlated with US Stocks. Gold Price is surprisingly uncorrelated with majorities. VIX shows a negative relationship with others.

\begin{figure}[htb]
  \centering
  \includegraphics[width=0.7\textwidth]{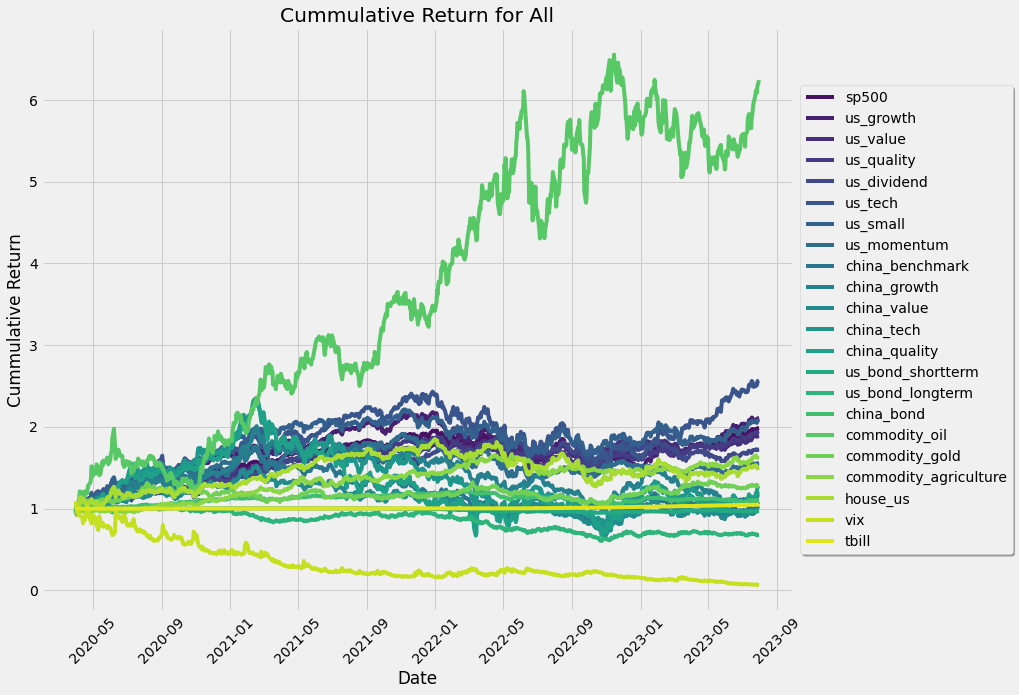}
  \caption{Cumulative Returns for Each Factor}
  \label{fig:cum_return}
\end{figure}

All the time series data are retrieved from Yahoo!Finance. Due to the availability of data (The API is under maintenance during my paper writing), the market cap data are manually collected from Yahoo!Finance, as Table \ref{tab:market_cap_weights}.

\begin{table}[H] 
\centering
\caption{Market Cap Weights}
\label{tab:market_cap_weights}
\begin{tabular}{lr}
\toprule
Asset Class & Market Cap Weight (Billion US Dollars)\\
\midrule
us\_growth & 0.279375 \\
us\_value & 0.230441 \\
us\_quality & 0.008476 \\
us\_dividend & 0.005083 \\
us\_tech & 0.019940 \\
us\_small & 0.192782 \\
us\_momentum & 0.014388 \\
china\_benchmark & 0.011964 \\
china\_growth & 0.000094 \\
china\_value & 0.007976 \\
china\_tech & 0.001290 \\
china\_quality & 0.000277 \\
us\_bond\_shortterm & 0.022551 \\
us\_bond\_longterm & 0.062102 \\
china\_bond & 0.000062 \\
commodity\_oil & 0.000963 \\
commodity\_gold & 0.043132 \\
commodity\_agriculture & 0.001313 \\
house\_us & 0.097431 \\
vix & 0.000360 \\
\bottomrule
\end{tabular}
\end{table}

\begin{sidewaystable}[htb]
\centering
\begin{threeparttable}
\caption{Summary Statistics}
\label{tab:sum_statistics}
\small 
\begin{tabular}{l *{8}{S[table-format=-1.6]}}

\toprule
& {count} & {mean} & {std} & {min} & {25\%} & {50\%} & {75\%} & {max} \\
\midrule
sp500 & 838.000000 & 0.000819 & 0.012185 & -0.057649 & -0.005953 & 0.000906 & 0.008093 & 0.067166 \\
us\_growth & 838.000000 & 0.000890 & 0.015485 & -0.052854 & -0.007666 & 0.001342 & 0.010127 & 0.078915 \\
us\_value & 838.000000 & 0.000759 & 0.011073 & -0.064920 & -0.004953 & 0.000864 & 0.007131 & 0.059691 \\
us\_quality & 838.000000 & 0.000795 & 0.012107 & -0.055998 & -0.005920 & 0.000894 & 0.007717 & 0.065513 \\
us\_dividend & 838.000000 & 0.000650 & 0.012504 & -0.069226 & -0.006355 & 0.000882 & 0.007129 & 0.065328 \\
us\_tech & 838.000000 & 0.001125 & 0.017552 & -0.059444 & -0.009032 & 0.001475 & 0.011383 & 0.088432 \\
us\_small & 838.000000 & 0.000879 & 0.015252 & -0.069041 & -0.008038 & 0.001094 & 0.010357 & 0.081969 \\
us\_momentum & 838.000000 & 0.000528 & 0.013960 & -0.050262 & -0.006901 & 0.001014 & 0.008715 & 0.071527 \\
china\_benchmark & 838.000000 & 0.000095 & 0.020208 & -0.096491 & -0.010557 & -0.000753 & 0.011093 & 0.209374 \\
china\_growth & 838.000000 & 0.000224 & 0.018457 & -0.079241 & -0.009940 & 0.000000 & 0.010440 & 0.162052 \\
china\_value & 838.000000 & 0.000038 & 0.020540 & -0.099916 & -0.011233 & -0.000786 & 0.010532 & 0.212412 \\
china\_tech & 838.000000 & 0.000193 & 0.024356 & -0.094042 & -0.013551 & -0.000570 & 0.013026 & 0.256760 \\
china\_quality & 838.000000 & 0.000296 & 0.031075 & -0.144995 & -0.016536 & -0.000921 & 0.016461 & 0.333933 \\
us\_bond\_shortterm & 838.000000 & -0.000038 & 0.001184 & -0.005299 & -0.000395 & 0.000000 & 0.000387 & 0.009954 \\
us\_bond\_longterm & 838.000000 & -0.000464 & 0.010393 & -0.034182 & -0.007624 & -0.000767 & 0.006142 & 0.038474 \\
china\_bond & 838.000000 & 0.000119 & 0.003574 & -0.014088 & -0.002028 & 0.000000 & 0.002147 & 0.017383 \\
commodity\_oil & 838.000000 & 0.002186 & 0.026586 & -0.097807 & -0.013820 & 0.001388 & 0.017624 & 0.159953 \\
commodity\_gold & 838.000000 & 0.000297 & 0.009579 & -0.053340 & -0.004664 & 0.000580 & 0.005802 & 0.030685 \\
commodity\_agriculture & 838.000000 & 0.000574 & 0.008719 & -0.038608 & -0.004789 & 0.000496 & 0.005709 & 0.029757 \\
house\_us & 838.000000 & 0.000486 & 0.014504 & -0.065712 & -0.007176 & 0.000461 & 0.008468 & 0.074074 \\
vix & 838.000000 & -0.003218 & 0.044699 & -0.135302 & -0.029367 & -0.009970 & 0.016966 & 0.335815 \\
tbill & 838.000000 & 0.000057 & 0.000075 & 0.000000 & 0.000002 & 0.000005 & 0.000125 & 0.000204 \\
\bottomrule
\end{tabular}
\begin{tablenotes}[flushleft]
\small
\item Note: Table shows sample space is 838, including panel data of 22 time-series (20 factors, 1 benchmark, and 1 risk-free rate) from 1st of April 2020 to 1st of August 2023, daily.
\end{tablenotes}
\end{threeparttable}
\end{sidewaystable}

\subsection{Risk Aversion of the Market}
By applying Equation \ref{eq:lambda_mkt}, plugging the benchmark's (S\&P500) mean return excess of the risk-free rate and benchmark's variance. We would get the risk aversion derived from the market. In our case, 

$$\lambda_{mkt} = 2.566/2 = 1.283$$

\subsection{Different Types of Risk Aversion}

The assignment asks to calculate the Black-Litterman weights under three different scenarios. Those three scenarios represent three types of investors.

\begin{itemize}
    \item Near-Kelly Investors: $\lambda = 0.01/2 = 0.005$
    \item Average Investors: $\lambda = 2.24/2 = 1.12$
    \item Risk Averse Investors: $\lambda = 6/3=2$
\end{itemize}

Note that $\lambda_{mkt} = 2.566/2= 1.283$ derived from our empirical data is similar to the given one $\lambda = 2.24/2 = 1.12$.

Later, we would substitute each type of risk aversion into our model, and see how weights are allocated by our model.

\section{Results} \label{sec:results}

In this section, we would firstly show the \textbf{static} Black-Litterman model results under different risk aversion scenarios. We would employ the market cap weights as our prior, and apply fixed views the same as the ones we introduced before. The second sub-section would present results applying \textbf{dynamic} prior, and back-testing results are also given. In the second subsection, we would split our data into different rolling time spans, and the prior would be replaced by the Markowitz weights during those time spans. In the meantime, views would also be periodically updated by a Deep Learning (LSTM) algorithm.

\subsection{Black-Litterman Model Results under Empirical Market Aversion}

As the Black-Litterman Model needs an Implied Equilibrium Return (prior) being inputted as a "benchmark", we would follow the instruction by \citet{heLitterman2002intuition}, using the market cap weights as our prior. Admittedly, there are inevitable drawbacks to using market cap weights in our case. For instance, those weights cannot be periodically updated due to the current data availability problem in Yahoo!Finance. We would still follow the instruction. Even though weights cannot be updated periodically, we could still obtain static results from the Black-Litterman model and make comparisons between results under different risk aversion scenarios. Meanwhile, we would plug in fixed views (Equation \ref{eq:views}) in each scenario.

\subsubsection{Results under Empirical Market Risk Aversion}

Table \ref{tab:pc_weights} shows how weights are allocated by using different weight allocation schemes. Table \ref{tab:pc_all} gives a comparison between the Black-Litterman weights/results and the Implied Equilibrium weights/returns. The "weights difference" column represents also the \textbf{"active weights"}. Table \ref{tab:all_BL_weights} shows the overall comparison between Black-Litterman model weights under different risk aversion scenarios.

\subsubsection{Results under Different Risk Aversion Scenario}

\begin{sidewaystable}[htb]
\centering
\begin{threeparttable}
\caption{Portfolio Weights by using different Weights Allocation Schemes}
\label{tab:pc_weights}
\tiny 
\begin{tabular}{l *{7}{S[table-format=2.1]}}
\hline
& {Market Cap Weights} & {Equal Weights} & {Implied Beta Weights} & {GMV Weights} & {Markowitz Weights with $\lambda=2$} & {Max Sharpe Weights}& {Black-Litterman Weights} \\
& {(\%)} & {(\%)} & {(\%)} & {(\%)} & {(\%)} & {(\%)}& {(\%)} \\
\hline
us\_growth & 27.9 & 5.0 & 46.9 & 0.0 & 0.0 & 0.0 & 33.3 \\
us\_value & 23.0 & 5.0 & 48.0 & 0.8 & 0.0 & 0.0 & 17.6 \\
us\_quality & 0.9 & 5.0 & 3.4 & 0.0 & 0.0 & 0.0 & 0.9 \\
us\_dividend & 0.5 & 5.0 & 2.9 & 0.0 & 0.0 & 0.0 & 0.5 \\
us\_tech & 2.0 & 5.0 & 2.4 & 0.0 & 3.4 & 18.1 & 2.0 \\
us\_small & 19.3 & 5.0 & -3.9 & 0.0 & 0.0 & 0.0 & 19.3 \\
us\_momentum & 1.4 & 5.0 & 0.8 & 0.7 & 0.0 & 0.0 & 7.1 \\
china\_benchmark & 1.2 & 5.0 & 1.1 & 0.0 & 0.0 & 0.0 & 0.8 \\
china\_growth & 0.0 & 5.0 & -0.3 & 0.0 & 0.0 & 0.0 & -0.4 \\
china\_value & 0.8 & 5.0 & -0.5 & 0.0 & 0.0 & 0.0 & 0.4 \\
china\_tech & 0.1 & 5.0 & -0.0 & 0.0 & 0.0 & 0.0 & -0.3 \\
china\_quality & 0.0 & 5.0 & -0.5 & 0.0 & 0.0 & 0.0 & -0.4 \\
us\_bond\_shortterm & 2.3 & 5.0 & -2.7 & 92.7 & 69.7 & 0.0 & 2.3 \\
us\_bond\_longterm & 6.2 & 5.0 & -0.0 & 0.0 & 0.0 & 0.0 & 4.3 \\
china\_bond & 0.0 & 5.0 & -0.4 & 3.7 & 13.3 & 18.9 & 0.0 \\
commodity\_oil & 0.1 & 5.0 & -0.1 & 0.3 & 3.8 & 17.1 & 0.1 \\
commodity\_gold & 4.3 & 5.0 & 0.1 & 0.0 & 0.0 & 8.3 & 4.3 \\
commodity\_agriculture & 0.1 & 5.0 & -0.0 & 1.5 & 9.8 & 37.6 & 0.1 \\
house\_us & 9.7 & 5.0 & -0.1 & 0.0 & 0.0 & 0.0 & 9.7 \\
vix & 0.0 & 5.0 & -0.2 & 0.4 & 0.0 & 0.0 & 0.0 \\
\hline
\end{tabular}
\begin{tablenotes}[flushleft]
\small
\item Note: Table shows weights derived from different weights allocation schemes introduced from the methodology section.
\end{tablenotes}
\end{threeparttable}
\end{sidewaystable}
\begin{sidewaystable}[htb]
\centering
\begin{threeparttable}
\caption{Portfolio Construction}
\label{tab:pc_all}
\small 
\begin{tabular}{lrrrrrr}
\toprule
 & Black-Litterman Return & pi & Return Difference & Black-Litterman Weights & Market Cap Weights & Weights Difference \\
\midrule
us\_growth & 0.14\% & -0.38\% & 0.52\% & 33.34\% & 27.94\% & 5.40\% \\
us\_value & 0.04\% & 0.56\% & -0.53\% & 17.64\% & 23.04\% & -5.40\% \\
us\_quality & -0.03\% & -0.01\% & -0.01\% & 0.85\% & 0.85\% & 0.00\% \\
us\_dividend & 0.30\% & 1.12\% & -0.81\% & 0.51\% & 0.51\% & 0.00\% \\
us\_tech & -0.02\% & -0.75\% & 0.74\% & 1.99\% & 1.99\% & -0.00\% \\
us\_small & 1.05\% & 1.16\% & -0.11\% & 19.28\% & 19.28\% & 0.00\% \\
us\_momentum & 0.59\% & 0.11\% & 0.48\% & 7.06\% & 1.44\% & 5.62\% \\
china\_benchmark & 1.49\% & 2.13\% & -0.63\% & 0.81\% & 1.20\% & -0.39\% \\
china\_growth & 1.49\% & 2.16\% & -0.67\% & -0.38\% & 0.01\% & -0.39\% \\
china\_value & 1.32\% & 2.08\% & -0.75\% & 0.41\% & 0.80\% & -0.39\% \\
china\_tech & 1.70\% & 2.25\% & -0.55\% & -0.26\% & 0.13\% & -0.39\% \\
china\_quality & 2.17\% & 2.48\% & -0.31\% & -0.36\% & 0.03\% & -0.39\% \\
us\_bond\_shortterm & 1.20\% & 2.04\% & -0.84\% & 2.26\% & 2.26\% & 0.00\% \\
us\_bond\_longterm & 2.19\% & 3.16\% & -0.97\% & 4.27\% & 6.21\% & -1.94\% \\
china\_bond & 1.24\% & 2.05\% & -0.81\% & 0.01\% & 0.01\% & -0.00\% \\
commodity\_oil & 0.06\% & 0.80\% & -0.74\% & 0.10\% & 0.10\% & 0.00\% \\
commodity\_gold & 1.76\% & 2.54\% & -0.78\% & 4.31\% & 4.31\% & -0.00\% \\
commodity\_agriculture & 1.09\% & 1.85\% & -0.75\% & 0.13\% & 0.13\% & -0.00\% \\
house\_us & 1.25\% & 1.67\% & -0.41\% & 9.74\% & 9.74\% & -0.00\% \\
vix & 3.68\% & 6.92\% & -3.24\% & 0.04\% & 0.04\% & -0.00\% \\
\bottomrule
\end{tabular}
\begin{tablenotes}[flushleft]
\small
\item Note: Table compares the benchmark (we use the market cap weights as our benchmark) and the Black-Litterman returns/weights. The risk-aversion passed into Black-Litterman model is the market level risk-aversion.
\end{tablenotes}
\end{threeparttable}
\end{sidewaystable}
\begin{sidewaystable}[htb]
\centering
\begin{threeparttable}
\caption{Weights and Returns of the Portfolio using Kelly Risk Aversion}
\label{tab:pc_kelly}
\small 
\begin{tabular}{lrrrrrr}
\toprule
& Black-Litterman Return & $\pi$ & Return Difference & Black-Litterman Weights & Market Cap Weights & Weights Difference \\
\midrule
us\_growth & 0.14\% & -0.38\% & 0.52\% & 17,107.41\% & 27.94\% & 17,079.47\% \\
us\_value & 0.04\% & 0.56\% & -0.53\% & 9,053.11\% & 23.04\% & 9,030.06\% \\
us\_quality & -0.03\% & -0.01\% & -0.01\% & 434.95\% & 0.85\% & 434.10\% \\
us\_dividend & 0.30\% & 1.12\% & -0.81\% & 260.81\% & 0.51\% & 260.30\% \\
us\_tech & -0.02\% & -0.75\% & 0.74\% & 1,023.18\% & 1.99\% & 1,021.19\% \\
us\_small & 1.05\% & 1.16\% & -0.11\% & 9,892.35\% & 19.28\% & 9,873.07\% \\
us\_momentum & 0.59\% & 0.11\% & 0.48\% & 3,621.10\% & 1.44\% & 3,619.66\% \\
china\_benchmark & 1.49\% & 2.13\% & -0.63\% & 414.48\% & 1.20\% & 413.28\% \\
china\_growth & 1.49\% & 2.16\% & -0.67\% & -194.63\% & 0.01\% & -194.64\% \\
china\_value & 1.32\% & 2.08\% & -0.75\% & 209.84\% & 0.80\% & 209.04\% \\
china\_tech & 1.70\% & 2.25\% & -0.55\% & -133.21\% & 0.13\% & -133.34\% \\
china\_quality & 2.17\% & 2.48\% & -0.31\% & -185.22\% & 0.03\% & -185.25\% \\
us\_bond\_shortterm & 1.20\% & 2.04\% & -0.84\% & 1,157.20\% & 2.26\% & 1,154.94\% \\
us\_bond\_longterm & 2.19\% & 3.16\% & -0.97\% & 2,189.55\% & 6.21\% & 2,183.34\% \\
china\_bond & 1.24\% & 2.05\% & -0.81\% & 3.16\% & 0.01\% & 3.15\% \\
commodity\_oil & 0.06\% & 0.80\% & -0.74\% & 49.40\% & 0.10\% & 49.31\% \\
commodity\_gold & 1.76\% & 2.54\% & -0.78\% & 2,213.28\% & 4.31\% & 2,208.97\% \\
commodity\_agriculture & 1.09\% & 1.85\% & -0.75\% & 67.39\% & 0.13\% & 67.26\% \\
house\_us & 1.25\% & 1.67\% & -0.41\% & 4,999.54\% & 9.74\% & 4,989.80\% \\
vix & 3.68\% & 6.92\% & -3.24\% & 18.47\% & 0.04\% & 18.44\% \\
\bottomrule
\end{tabular}
\begin{tablenotes}[flushleft]
\small
\item Note: Weights and Returns of the Portfolio using Kelly Risk Aversion passed into the Black-Litterman Model. Kelly risk aversion sets $\lambda = 0.01/2 = 0.005$. From the table, we can find that Black-Litterman weights are enlarged. In other words, investors are less risk averse so that they would like to take more leverage.
\end{tablenotes}
\end{threeparttable}
\end{sidewaystable}

\begin{sidewaystable}[htb]
\centering
\begin{threeparttable}
\caption{Weights and Returns of the Portfolio using Market Level Risk Aversion}
\label{tab:pc_all_mkt}
\small 
\begin{tabular}{lrrrrrr}
\toprule
& Black-Litterman Return & $\pi$ & Return Difference & Black-Litterman Weights & Market Cap Weights & Weights Difference \\
\midrule
us\_growth & 0.14\% & -0.38\% & 0.52\% & 76.37\% & 27.94\% & 48.43\% \\
us\_value & 0.04\% & 0.56\% & -0.53\% & 40.42\% & 23.04\% & 17.37\% \\
us\_quality & -0.03\% & -0.01\% & -0.01\% & 1.94\% & 0.85\% & 1.09\% \\
us\_dividend & 0.30\% & 1.12\% & -0.81\% & 1.16\% & 0.51\% & 0.66\% \\
us\_tech & -0.02\% & -0.75\% & 0.74\% & 4.57\% & 1.99\% & 2.57\% \\
us\_small & 1.05\% & 1.16\% & -0.11\% & 44.16\% & 19.28\% & 24.88\% \\
us\_momentum & 0.59\% & 0.11\% & 0.48\% & 16.17\% & 1.44\% & 14.73\% \\
china\_benchmark & 1.49\% & 2.13\% & -0.63\% & 1.85\% & 1.20\% & 0.65\% \\
china\_growth & 1.49\% & 2.16\% & -0.67\% & -0.87\% & 0.01\% & -0.88\% \\
china\_value & 1.32\% & 2.08\% & -0.75\% & 0.94\% & 0.80\% & 0.14\% \\
china\_tech & 1.70\% & 2.25\% & -0.55\% & -0.59\% & 0.13\% & -0.72\% \\
china\_quality & 2.17\% & 2.48\% & -0.31\% & -0.83\% & 0.03\% & -0.85\% \\
us\_bond\_shortterm & 1.20\% & 2.04\% & -0.84\% & 5.17\% & 2.26\% & 2.91\% \\
us\_bond\_longterm & 2.19\% & 3.16\% & -0.97\% & 9.77\% & 6.21\% & 3.56\% \\
china\_bond & 1.24\% & 2.05\% & -0.81\% & 0.01\% & 0.01\% & 0.01\% \\
commodity\_oil & 0.06\% & 0.80\% & -0.74\% & 0.22\% & 0.10\% & 0.12\% \\
commodity\_gold & 1.76\% & 2.54\% & -0.78\% & 9.88\% & 4.31\% & 5.57\% \\
commodity\_agriculture & 1.09\% & 1.85\% & -0.75\% & 0.30\% & 0.13\% & 0.17\% \\
house\_us & 1.25\% & 1.67\% & -0.41\% & 22.32\% & 9.74\% & 12.58\% \\
vix & 3.68\% & 6.92\% & -3.24\% & 0.08\% & 0.04\% & 0.05\% \\
\bottomrule
\end{tabular}
\begin{tablenotes}[flushleft]
\small
\item Note: Weights and Returns of the Portfolio using market level Risk Aversion passed into the Black-Litterman Model. Kelly risk aversion sets $\lambda = 2.24/2 = 1.12$. From the table, we can find that Black-Litterman weights by using risk aversion at the market average level. In other words, investors are accept the same level of risk aversion with the overall market.
\end{tablenotes}
\end{threeparttable}
\end{sidewaystable}

\begin{sidewaystable}[htb]
\centering
\begin{threeparttable}
\caption{Weights and Returns of the Portfolio using Risk Averse Risk Aversion}
\label{tab:pc_all_averse}
\small 
\begin{tabular}{lrrrrrr}
\toprule
& Black-Litterman Return & $\pi$ & Return Difference & Black-Litterman Weights & Market Cap Weights & Weights Difference \\
\midrule
us\_growth & 0.14\% & -0.38\% & 0.52\% & 28.51\% & 27.94\% & 0.57\% \\
us\_value & 0.04\% & 0.56\% & -0.53\% & 15.09\% & 23.04\% & -7.96\% \\
us\_quality & -0.03\% & -0.01\% & -0.01\% & 0.72\% & 0.85\% & -0.12\% \\
us\_dividend & 0.30\% & 1.12\% & -0.81\% & 0.43\% & 0.51\% & -0.07\% \\
us\_tech & -0.02\% & -0.75\% & 0.74\% & 1.71\% & 1.99\% & -0.29\% \\
us\_small & 1.05\% & 1.16\% & -0.11\% & 16.49\% & 19.28\% & -2.79\% \\
us\_momentum & 0.59\% & 0.11\% & 0.48\% & 6.04\% & 1.44\% & 4.60\% \\
china\_benchmark & 1.49\% & 2.13\% & -0.63\% & 0.69\% & 1.20\% & -0.51\% \\
china\_growth & 1.49\% & 2.16\% & -0.67\% & -0.32\% & 0.01\% & -0.33\% \\
china\_value & 1.32\% & 2.08\% & -0.75\% & 0.35\% & 0.80\% & -0.45\% \\
china\_tech & 1.70\% & 2.25\% & -0.55\% & -0.22\% & 0.13\% & -0.35\% \\
china\_quality & 2.17\% & 2.48\% & -0.31\% & -0.31\% & 0.03\% & -0.34\% \\
us\_bond\_shortterm & 1.20\% & 2.04\% & -0.84\% & 1.93\% & 2.26\% & -0.33\% \\
us\_bond\_longterm & 2.19\% & 3.16\% & -0.97\% & 3.65\% & 6.21\% & -2.56\% \\
china\_bond & 1.24\% & 2.05\% & -0.81\% & 0.01\% & 0.01\% & -0.00\% \\
commodity\_oil & 0.06\% & 0.80\% & -0.74\% & 0.08\% & 0.10\% & -0.01\% \\
commodity\_gold & 1.76\% & 2.54\% & -0.78\% & 3.69\% & 4.31\% & -0.62\% \\
commodity\_agriculture & 1.09\% & 1.85\% & -0.75\% & 0.11\% & 0.13\% & -0.02\% \\
house\_us & 1.25\% & 1.67\% & -0.41\% & 8.33\% & 9.74\% & -1.41\% \\
vix & 3.68\% & 6.92\% & -3.24\% & 0.03\% & 0.04\% & -0.01\% \\
\bottomrule
\end{tabular}
\begin{tablenotes}[flushleft]
\small
\item Note: Weights and Returns of the Portfolio using Risk Averse Risk Aversion passed into the Black-Litterman Model. Risk Averse risk aversion sets $\lambda = 6/2 = 3$. From the table, we can find that Black-Litterman weights are more conservative. In other words, investors are more risk averse so that they would like to take no risks.
\end{tablenotes}
\end{threeparttable}
\end{sidewaystable}

\begin{threeparttable}
\centering
\caption{Black-Litterman Results under Different Risk Aversion Scenarios}
\label{tab:all_BL_weights}
\begin{tabular}{lrrrr}
\toprule
 & Empirical Mkt & Kelly & Market & Risk Averse \\
\midrule
us\_growth & 33.34\% & 17,107.41\% & 76.37\% & 28.51\% \\
us\_value & 17.64\% & 9,053.11\% & 40.42\% & 15.09\% \\
us\_quality & 0.85\% & 434.95\% & 1.94\% & 0.72\% \\
us\_dividend & 0.51\% & 260.81\% & 1.16\% & 0.43\% \\
us\_tech & 1.99\% & 1,023.18\% & 4.57\% & 1.71\% \\
us\_small & 19.28\% & 9,892.35\% & 44.16\% & 16.49\% \\
us\_momentum & 7.06\% & 3,621.10\% & 16.17\% & 6.04\% \\
china\_benchmark & 0.81\% & 414.48\% & 1.85\% & 0.69\% \\
china\_growth & -0.38\% & -194.63\% & -0.87\% & -0.32\% \\
china\_value & 0.41\% & 209.84\% & 0.94\% & 0.35\% \\
china\_tech & -0.26\% & -133.21\% & -0.59\% & -0.22\% \\
china\_quality & -0.36\% & -185.22\% & -0.83\% & -0.31\% \\
us\_bond\_shortterm & 2.26\% & 1,157.20\% & 5.17\% & 1.93\% \\
us\_bond\_longterm & 4.27\% & 2,189.55\% & 9.77\% & 3.65\% \\
china\_bond & 0.01\% & 3.16\% & 0.01\% & 0.01\% \\
commodity\_oil & 0.10\% & 49.40\% & 0.22\% & 0.08\% \\
commodity\_gold & 4.31\% & 2,213.28\% & 9.88\% & 3.69\% \\
commodity\_agriculture & 0.13\% & 67.39\% & 0.30\% & 0.11\% \\
house\_us & 9.74\% & 4,999.54\% & 22.32\% & 8.33\% \\
vix & 0.04\% & 18.47\% & 0.08\% & 0.03\% \\
\bottomrule
\end{tabular}
\begin{tablenotes}[flushleft]
\small
\item Note: The first column uses the risk aversion derived from the empirical market risk aversion, $\lambda_{mkt} = 2.566/2=1.283$, the follow columns denote risk aversion be kelly $\lambda = 0.01/2=0.005$, Average Investors' $\lambda = 2.24/2 = 1.12$, and Risk Averse Investors $\lambda = 6/2=3$.
\end{tablenotes}
\end{threeparttable}

\begin{figure}[ht!]
  \centering
  \includegraphics[width=0.8\textwidth]{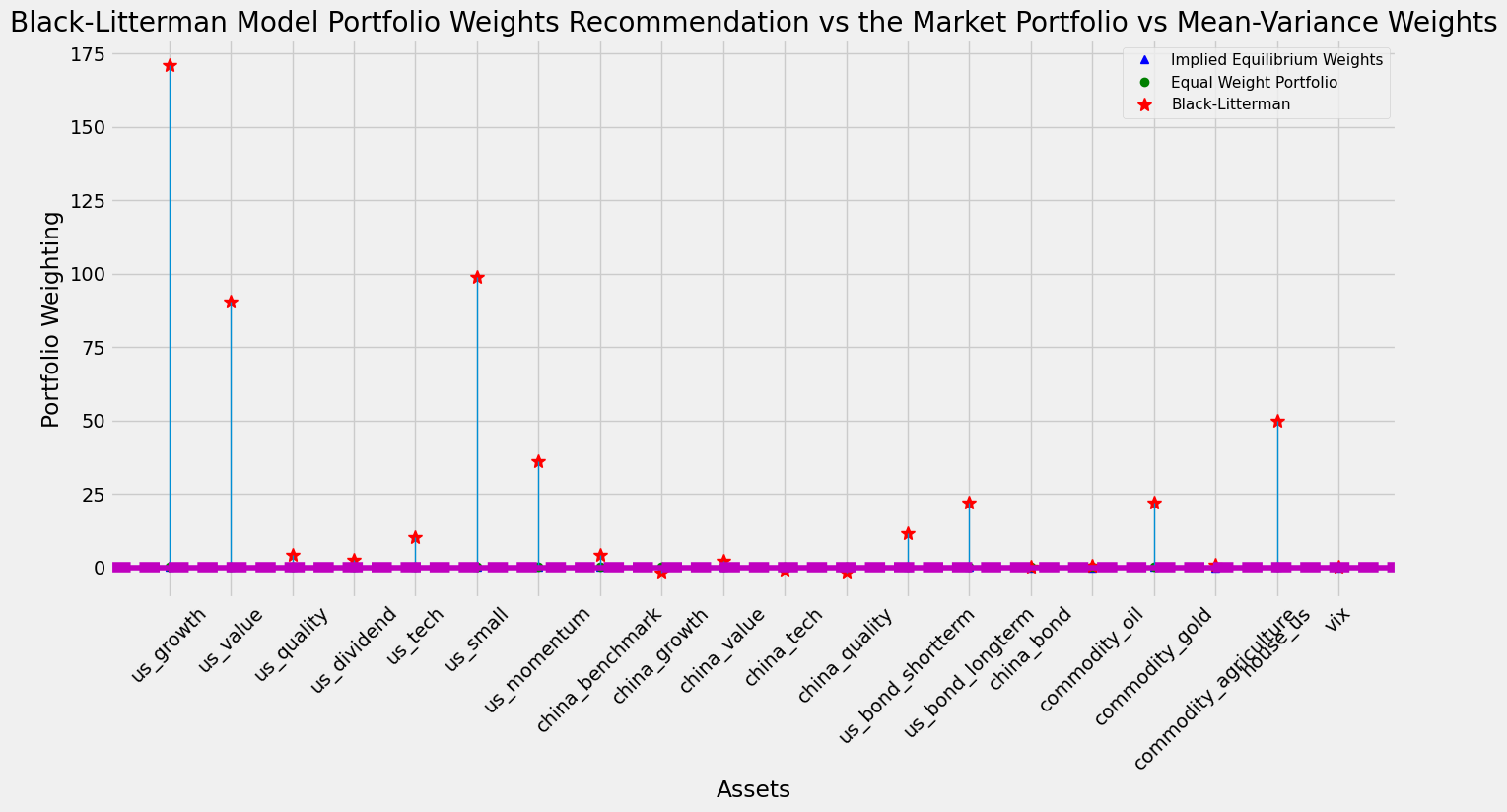}
  \caption{Black-Litterman Return \& Weights by using Near-Kelly Risk Aversion}
  \label{fig:plot_pc_kelly}
\end{figure}

\begin{figure}[ht!]
  \centering
  \includegraphics[width=0.8\textwidth]{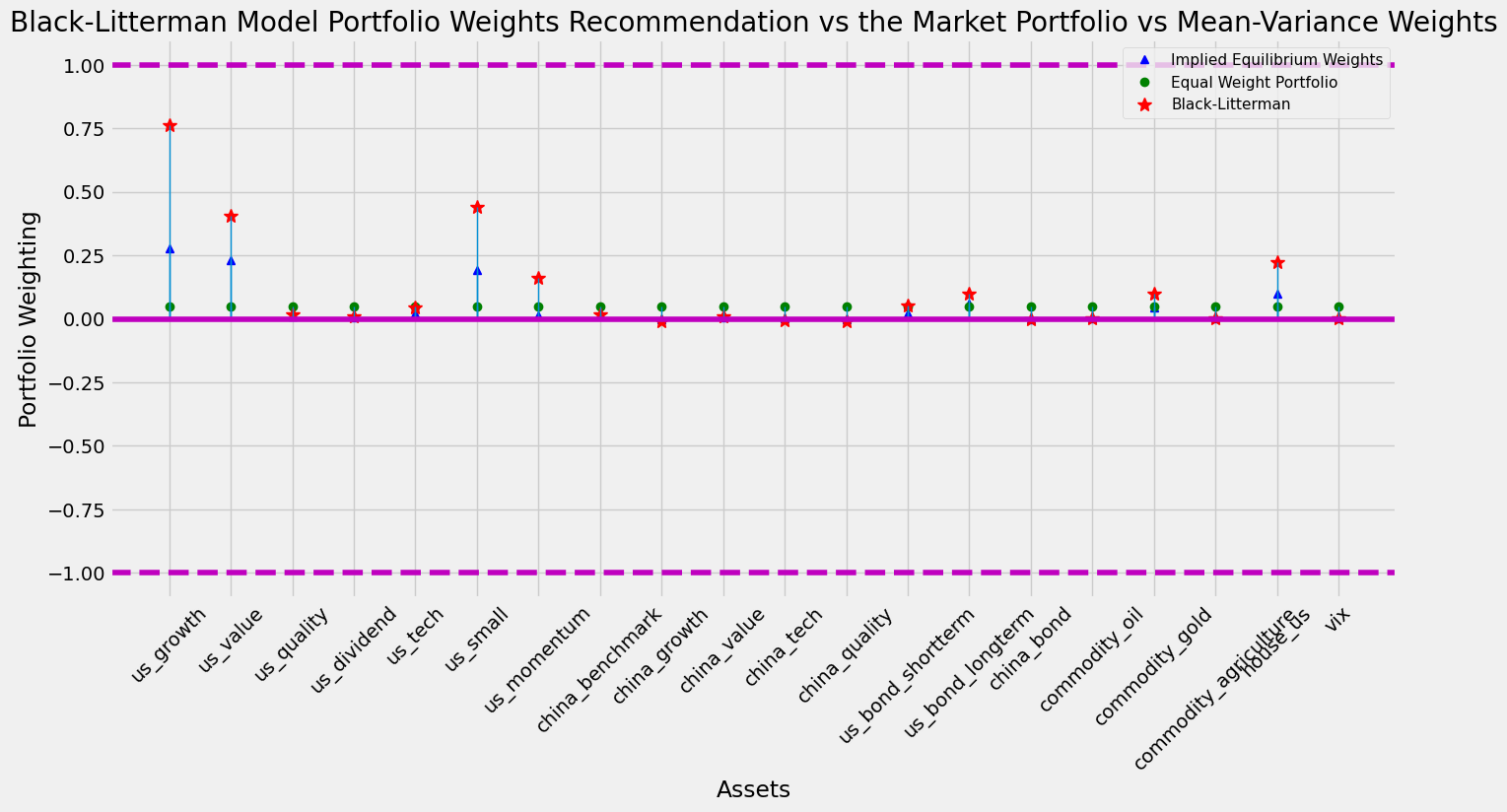}
  \caption{Black-Litterman Return \& Weights by using Average Investors Risk Aversion}
  \label{fig:plot_pc_mkt}
\end{figure}

\begin{figure}[ht!]
  \centering
  \includegraphics[width=0.8\textwidth]{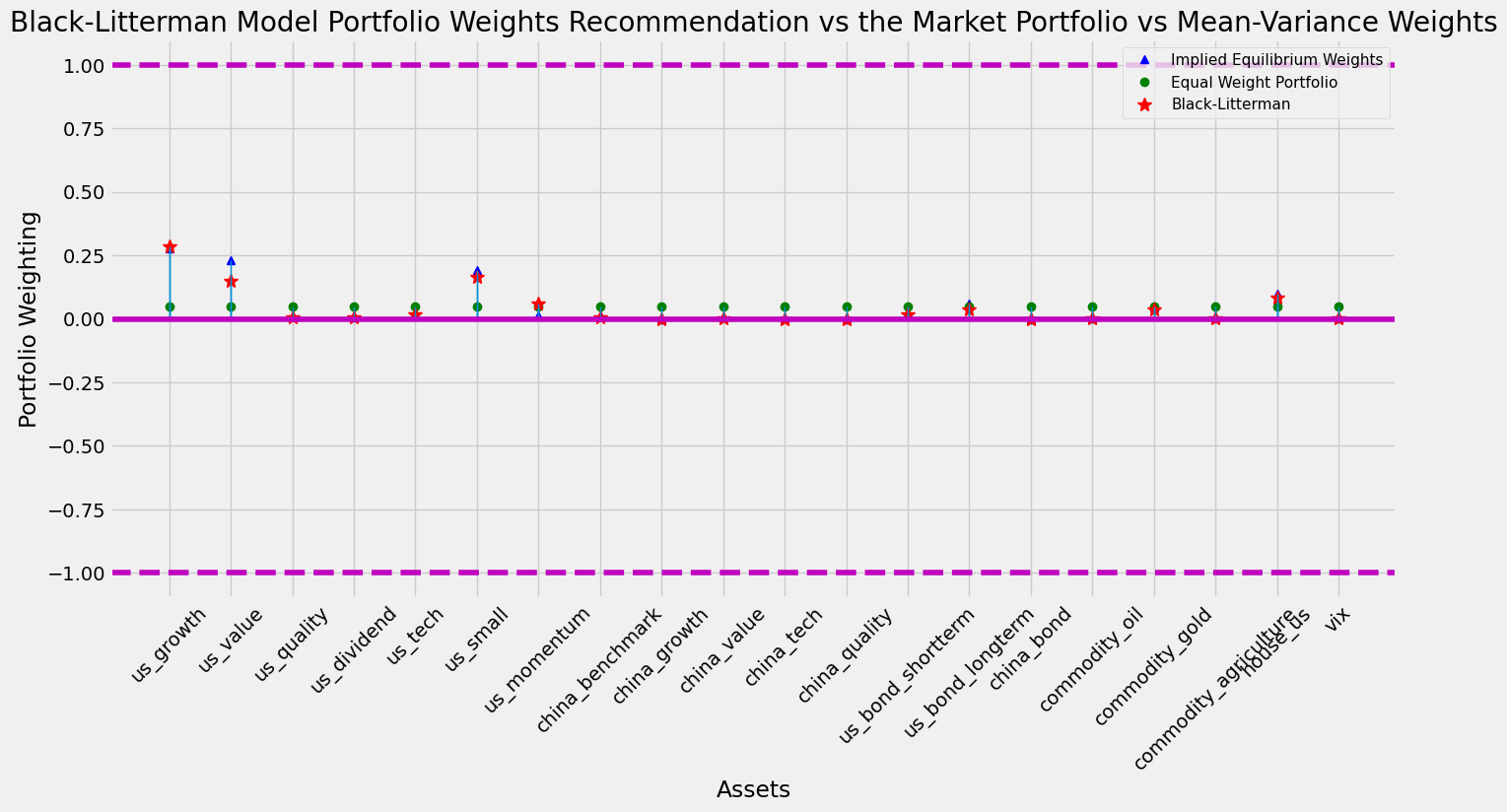}
  \caption{Black-Litterman Return \& Weights by using Risk Averse Risk Aversion}
  \label{fig:plot_pc_averse}
\end{figure}

Table \ref{tab:pc_kelly}, Table \ref{tab:pc_all_mkt}, and Table \ref{tab:pc_all_averse} shows the results under near-Kelly, average Investors, and risk averse risk aversion, respectively. Figure \ref{fig:plot_pc_kelly}, Figure \ref{fig:plot_pc_mkt}, and Figure \ref{fig:plot_pc_averse} show the weight allocation under near-Kelly, average investors, and risk averse risk aversion with respect to the prior, respectively. 

From those Tables and Figures, we can find that the Black-Litterman weights are enlarged under the near-Kelly risk aversion that sets $\lambda = 0.01/2 = 0.005$. Investors are less risk averse so they would like to take more leverage. In contrast, the Black-Litterman weights are more conservative, under the Risk Averse risk aversion setting to be $\lambda = 6/2 = 3$. In other words, investors are more risk averse so they would like to take no risks.

\begin{figure}[ht!]
  \centering
  \includegraphics[width=0.8\textwidth]{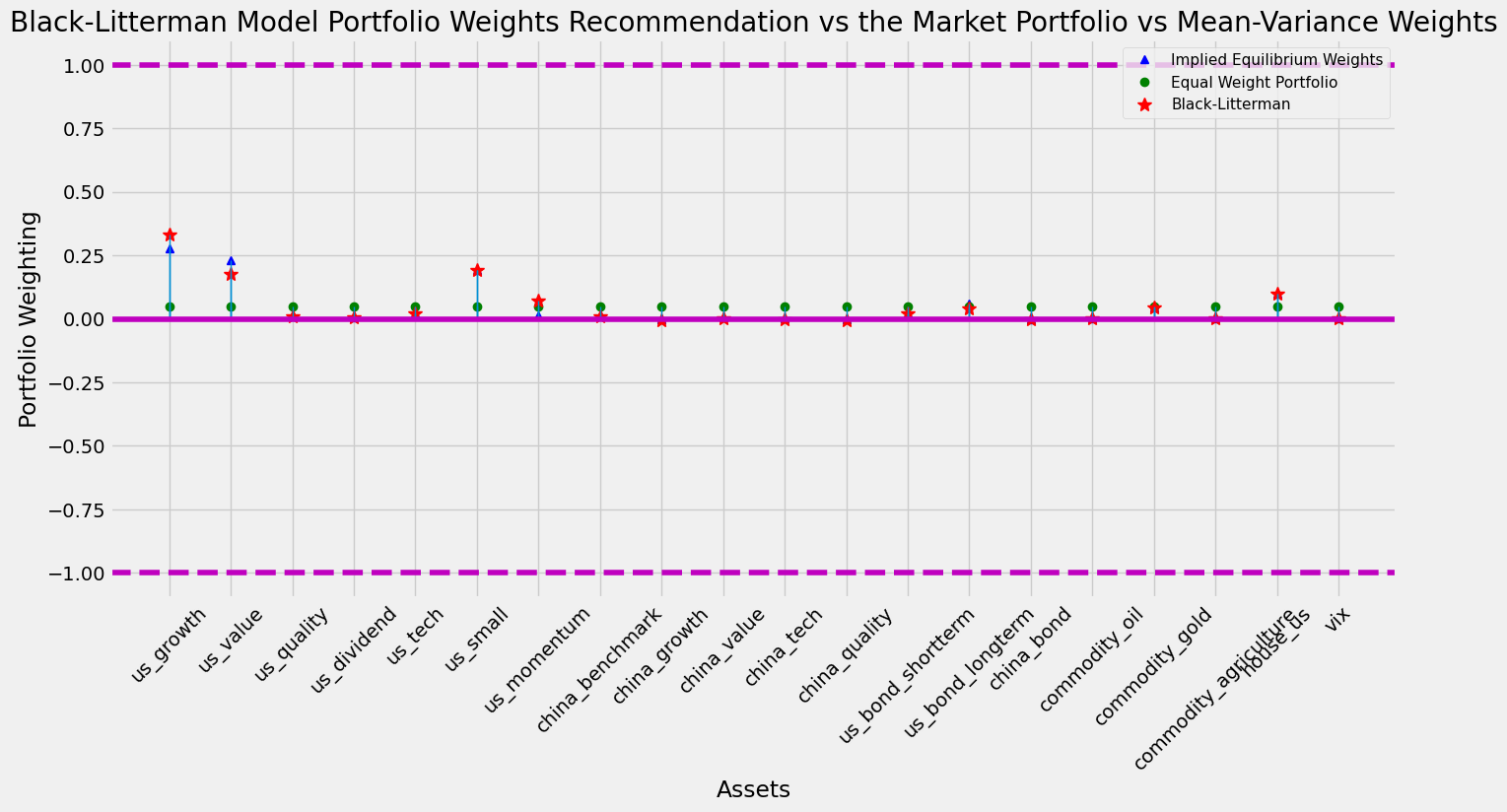}
  \caption{Black-Litterman Return \& Weights by using $lambda_{mkt}$}
  \label{fig:plot_pc}
\end{figure}

\begin{figure}[ht!]
  \centering
  \includegraphics[width=0.7\textwidth]{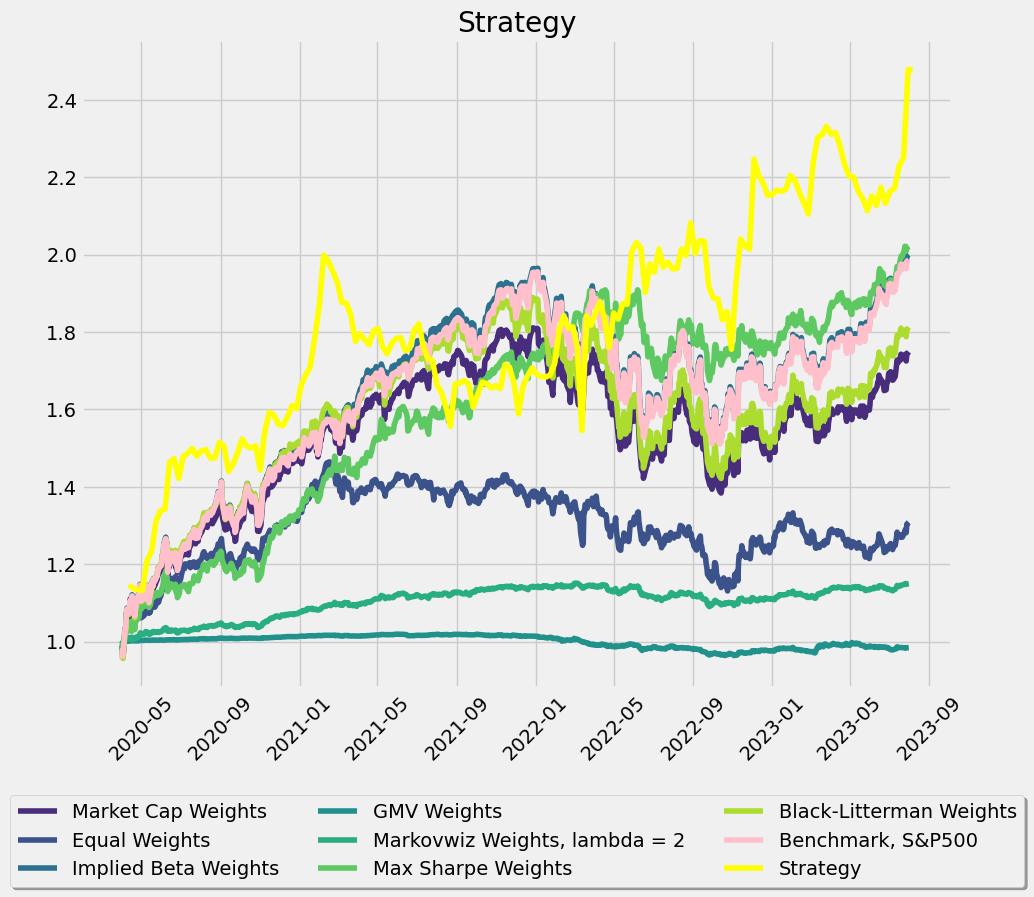}
  \caption{Static Weight Allocation Back-testing \& A Simple Contrarian Strategy}
  \label{fig:plot_momentum}
\end{figure}

\subsubsection{Applying the Static Weight Allocation and A Simply Back-testing using Contrarian Strategy}

Figure \ref{fig:plot_momentum} presents portfolio returns movement by using static weights employing different weight allocation schemes. We can find that the Black-Litterman Weights curve moves close to the prior (Market Cap Weights), with a little deviation. Note that the Black-Litterman model in this back-testing uses empirical market risk aversion.

P.S. The following paragraph is additional, we just apply a simple strategy to see if the back-testing framework works and see how the static weights allocation schemes change the cumulative returns of those factors. \textit{We also create a simple Contrarian strategy, which is captured by the Yellow curve. In the strategy, we resample our dataset on a weekly basis, choose 5 factors over 20 with the lowest cumulative returns over the week, and invest those 5 factors with equal weights for the next week. Repeat the process over and over. Strategy surprisingly performs not bad for most of the time.}

\subsection{Dynamic Black-Litterman Results \& Views from Deep Learning Algorithm}

\begin{figure}[ht!]
  \centering
  \includegraphics[width=0.7\textwidth]{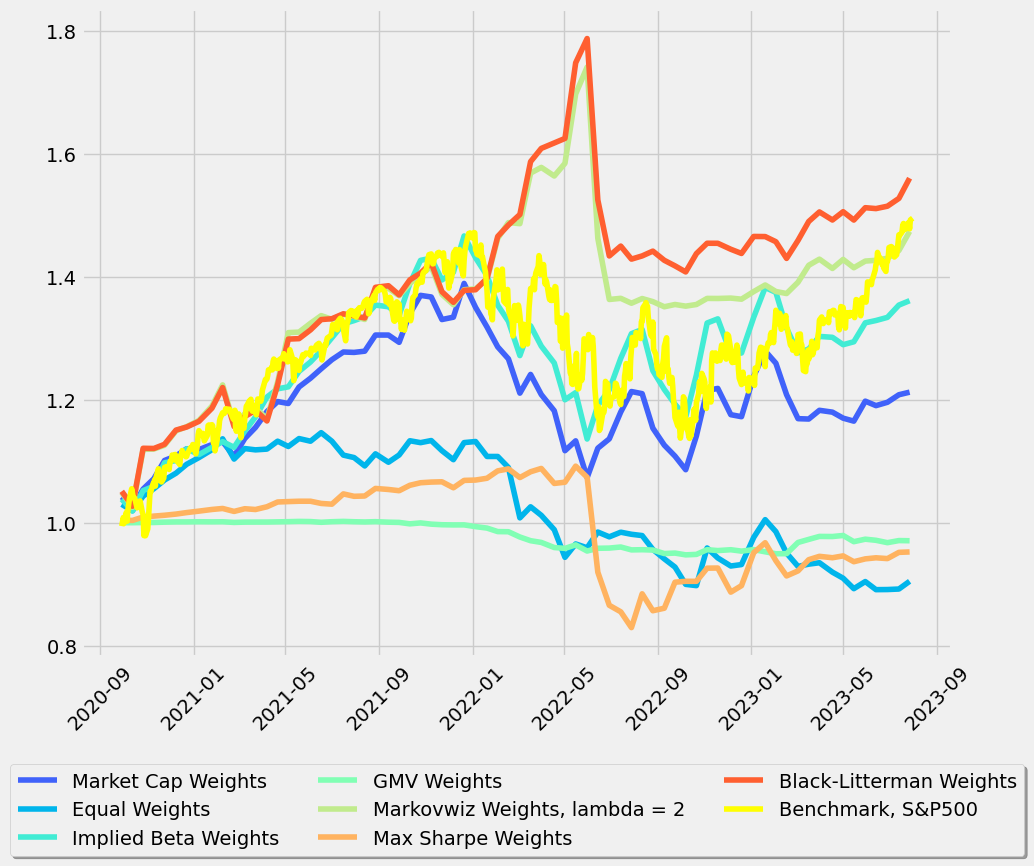}
  \caption{Dynamic Weight Allocation Back-testing \& Using LSTM to Generate Views}
  \label{fig:dynamic}
\end{figure}

In the previous subsection, we use a fixed prior to generating the static Black-Litterman results. In this subsection, we would produce a Deep Learning LSTM algorithm to update our views periodically. The methodology has been introduced in Section \ref{sec:LSTM}. 

We note that the prior of the Black-Litterman Model is replaced by the Markowitz mean-variance weights with $\lambda = 2$ because we need the dynamic re-balance of weights in this sub-section. The market cap weights are definitely applicable but are not employed as the prior here, because we currently do not have periodical market cap data. We just aim to show the model is working on a dynamic basis. One might debate the Markowitz mean-variance weights are not reliable and not stable. That is true, but our model is trying to split the sample into separate time spans, which would, to some degree, alleviate the problem that historical returns cannot well represent the true returns.

We also found that the red curve, the Black-Litterman cumulative returns outperforms others, especially during the first quarter of 2022 when most of the other weight allocation schemes generate negative returns. That out-performance is largely caused by optimistic Markowitz Weights results.

\section{Robustness Checking} \label{sec:robustness checking}

\subsection{Weight Allocation Movements over Time}

\begin{figure}[H]
  \centering
  \includegraphics[width=1\textwidth]{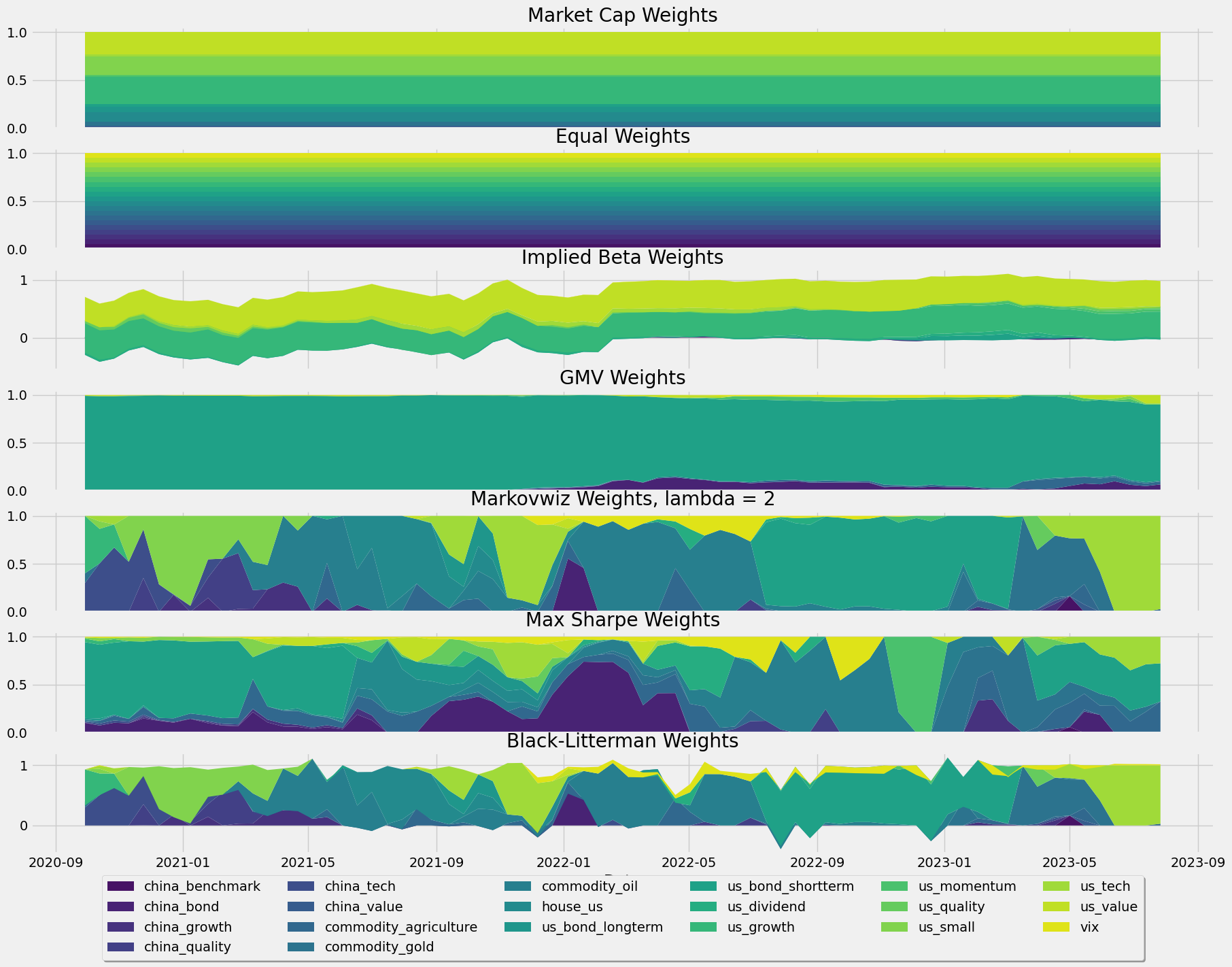}
  \caption{Dynamic Weight Allocation Moves Over Time}
  \label{fig:stack_plots}
\end{figure}

\begin{figure}[ht]
  \centering
  \includegraphics[width=1\textwidth]{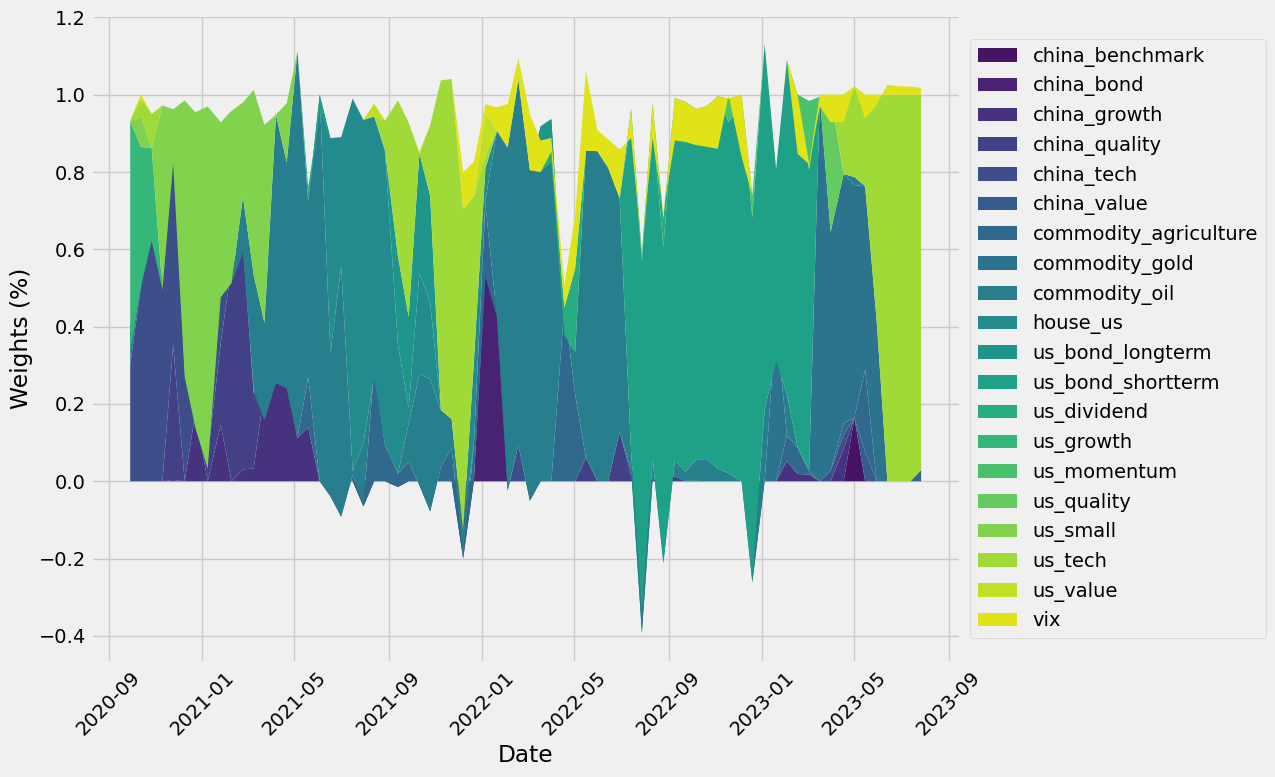}
  \caption{Dynamic Black-Litterman Weight Allocation Moves Over Time}
  \label{fig:stack_plots_BL_only}
\end{figure}

Let's now go through how weight allocation would change over time. Figure \ref{fig:stack_plots} shows how each scheme of weight allocation moves over time. We sadly found that there are corner solutions appearing in Global Minimum Variance (GMV) weight allocation. Long-term US bonds account for a large percentage of the portfolio over most of the time, probably due to the reason that the objective function of GMV is to choose less volatile factors. Figure \ref{fig:stack_plots_BL_only} demonstrates the Black-Litterman Weight Allocation movement only.

\begin{figure}[ht]
  \centering
  \includegraphics[width=1\textwidth]{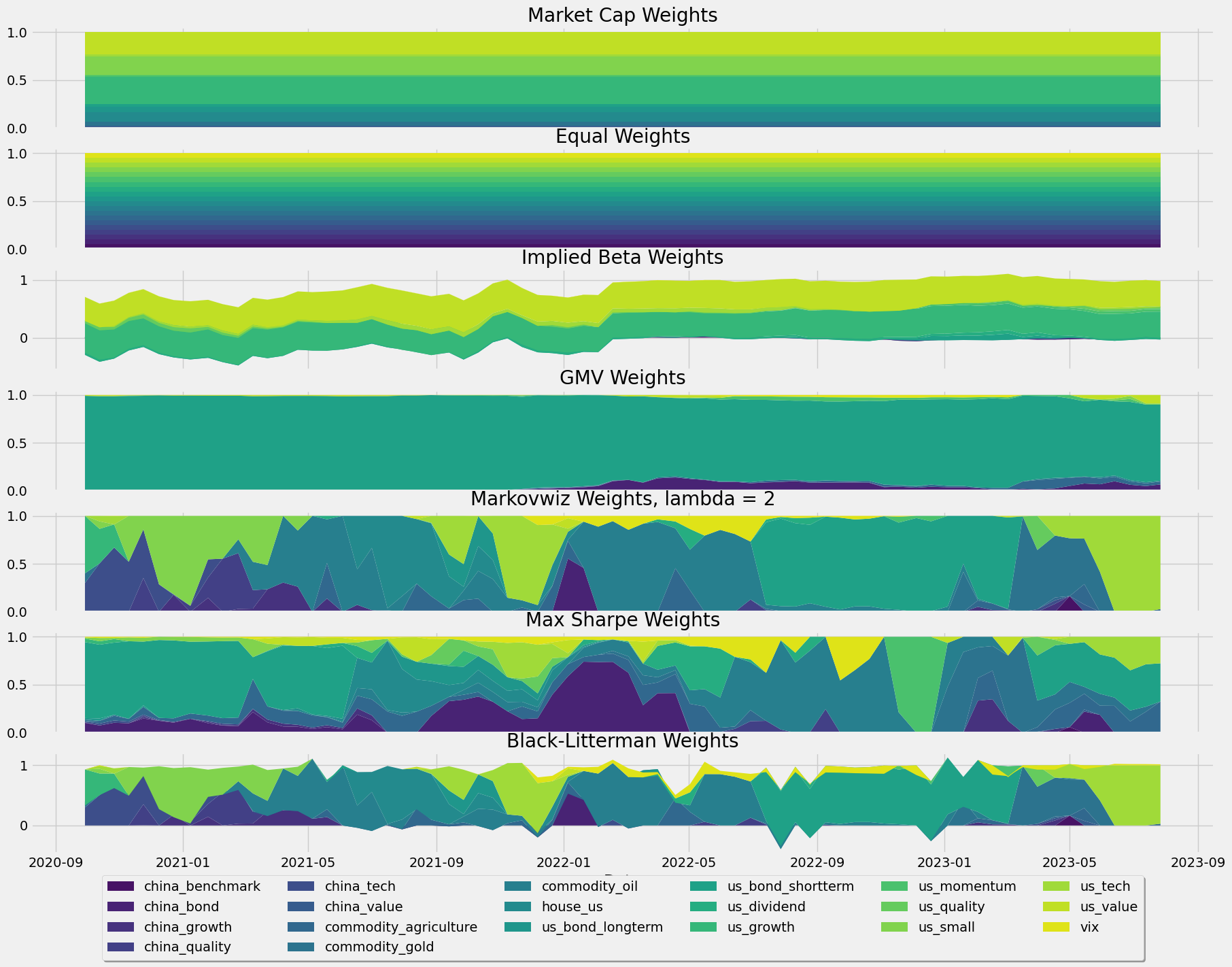}
  \caption{Dynamic Weight Allocation with Shrink Variance Moves Over Time}
  \label{fig:stack_plots_shrink}
\end{figure}

Meanwhile, we do also find that Markowitz's weights appear to have corner solutions over certain periods. As seriously considering the suggestion from the \textit{CQF Slides}, we use the LSTM model to generate views and use the Shrinkage Method \citep{ledoit2003improved} to improve the variance-covariance of our factors. Figure \ref{fig:stack_plots_shrink} shows the Weight Allocation for different schemes while using the Shrinkage Variance. We cannot found clear different between Figure \ref{fig:stack_plots} and Figure \ref{fig:stack_plots_shrink}
, but we could find the variance-covariance metrics (which could be found in the code) are indeed not the same for the two methods.

\subsection{Weight Allocation Movements w.r.t. Volatility Changes}

\begin{figure}[ht!]
  \centering
  \includegraphics[width=0.55\textwidth]{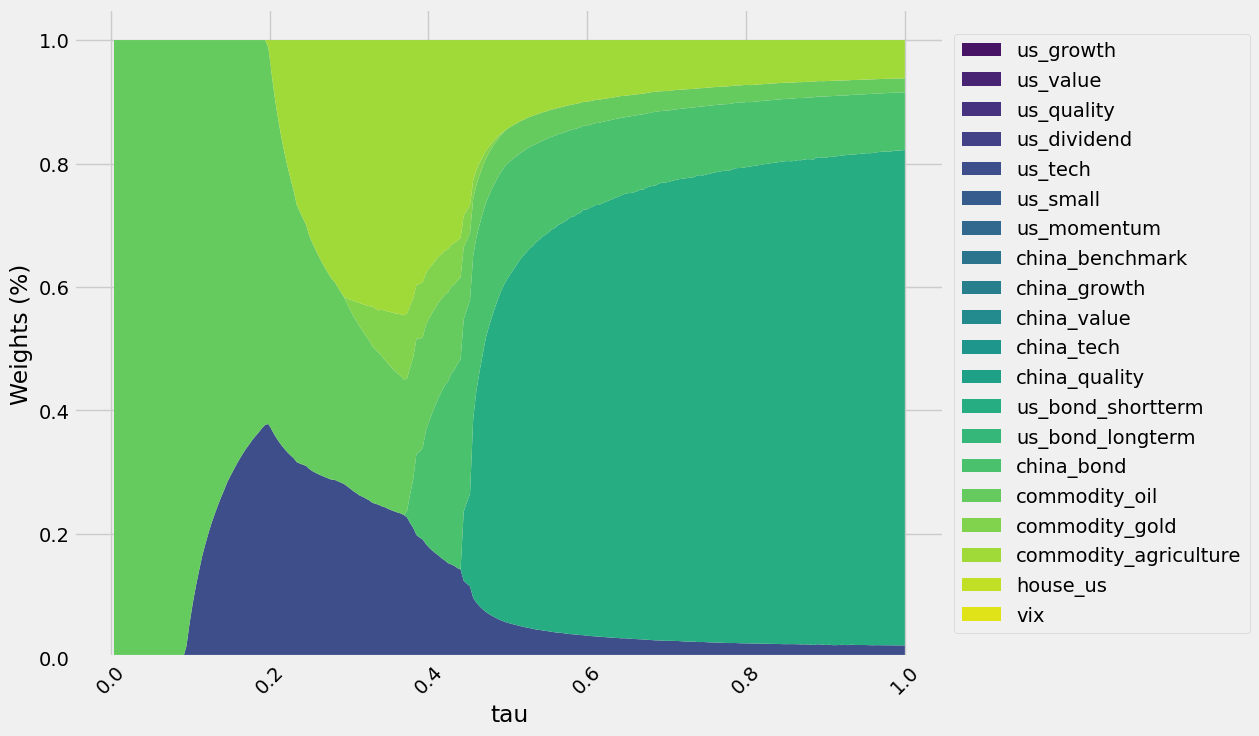}
  \caption{Prior Weights Move with Volatility Changes}
  \label{fig:vol_prior}
\end{figure}

\begin{figure}[ht!]
  \centering
  \includegraphics[width=0.55\textwidth]{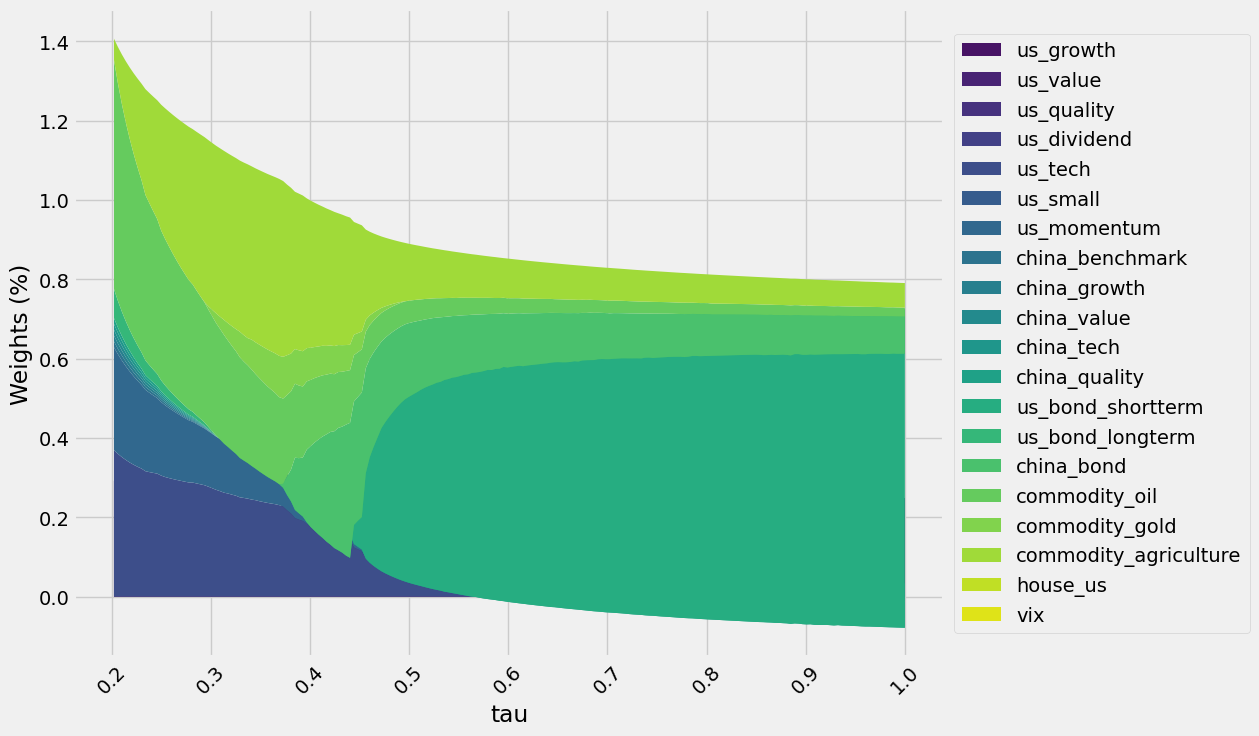}
  \caption{Posterior Weights Move with Volatility Changes}
  \label{fig:vol_posterior}
\end{figure}

\begin{figure}[ht!]
  \centering
  \includegraphics[width=0.55\textwidth]{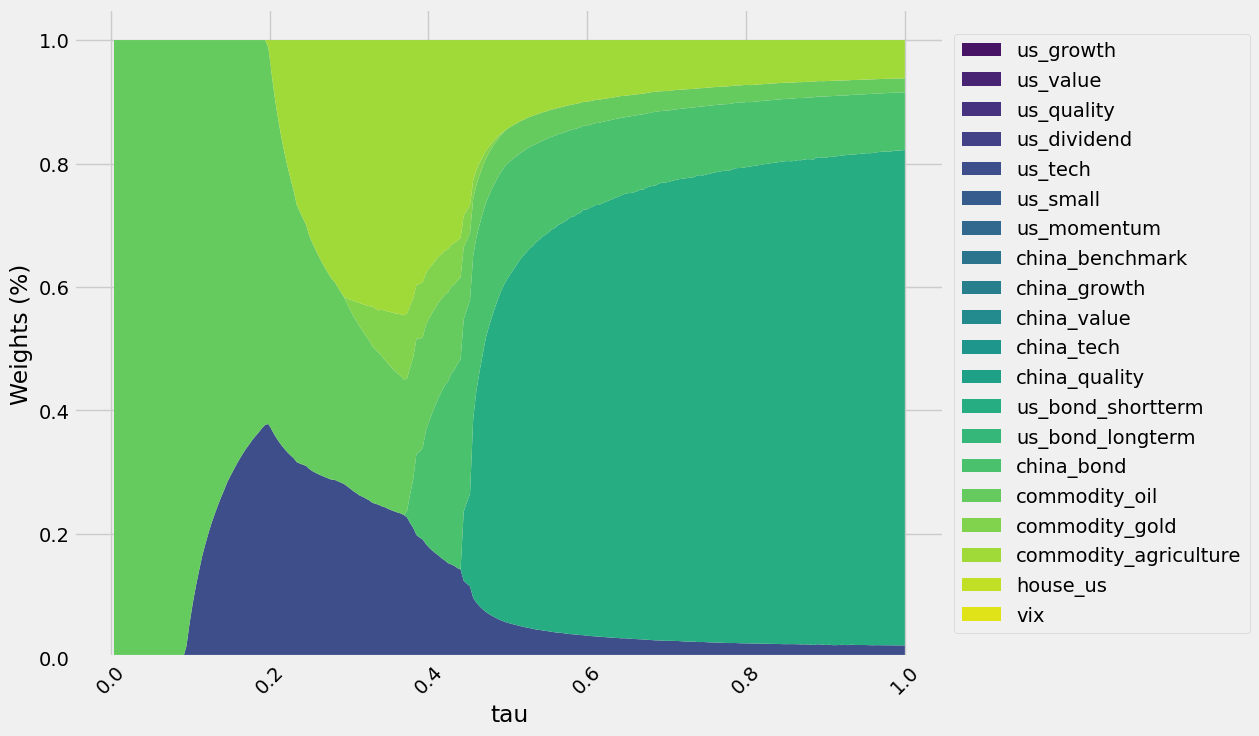}
  \caption{Prior Weights with Shrinkage Volatility Move with Volatility Changes}
  \label{fig:vol_prior_shrink}
\end{figure}

\begin{figure}[ht!]
  \centering
  \includegraphics[width=0.55\textwidth]{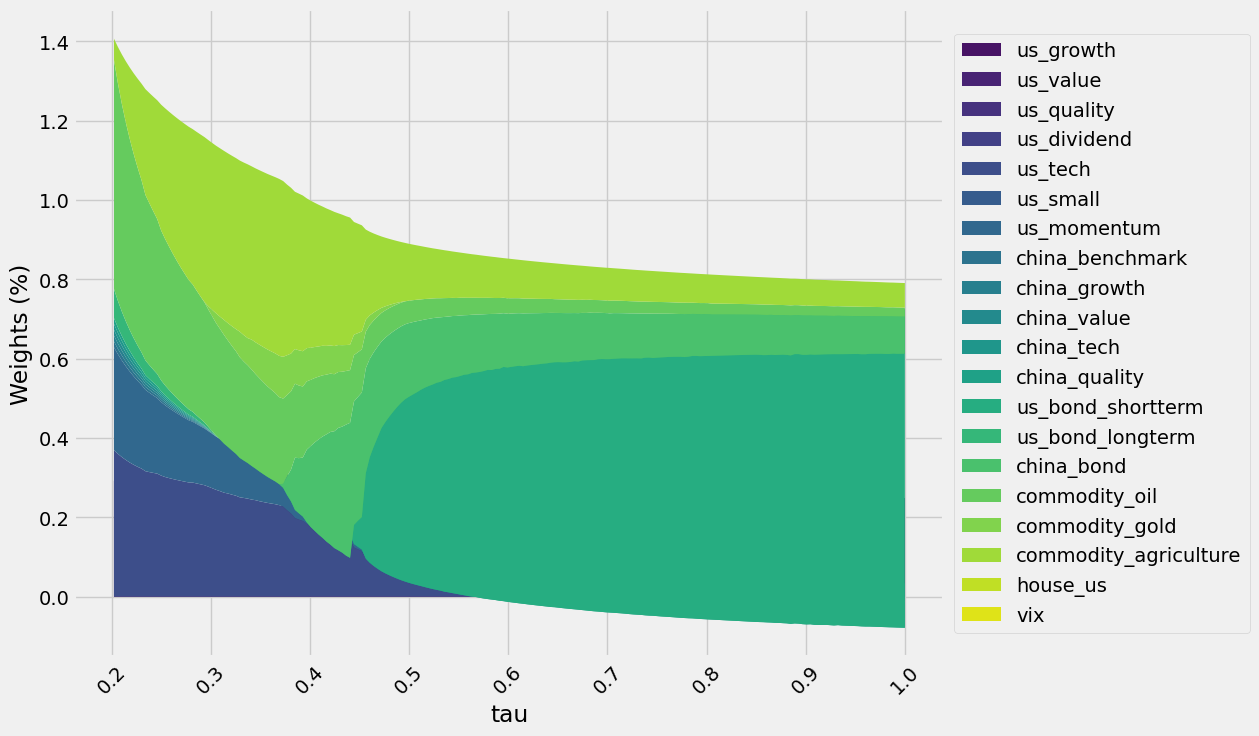}
  \caption{Posterior Weights with Shrinkage Volatility Move with Volatility Changes}
  \label{fig:vol_posterior_shrink}
\end{figure}

In this section, we would present how would the prior and the posterior evolve with the increase in volatility ($\Sigma_{input} = tau \times \Sigma$, the volatility $\Sigma_{input}$ increases with $\tau$ increase. Note that $tau \neq \tau$, as $\tau$ changes would not change the Black-Litterman results, while $tau$ is just a parameter used to enlarge volatility in this particular section), as Figure \ref{fig:vol_prior} and Figure \ref{fig:vol_posterior}. To improve the robustness and try to avoid the corner solutions, we apply the shrinkage variance to generate those weights, as Figure \ref{fig:vol_prior_shrink} and Figure \ref{fig:vol_posterior_shrink}. The difference between results using the basic variance and shrinkage variance cannot be clearly identified.

\section{Discussion and Improvement} \label{sec:Discussion}

Here below, we would share some discussion, including some ideas or drawbacks that cannot be covered or solved in this paper, and would suggest further improvements.

1. Historical Return and Variance Covariance might not be reliable. Although the shrinkage estimator for variance\citep{ledoit2003improved} is applied as the comparison, there seems to have little impact on the weight allocation.

2. The sample space we have chosen avoids the periods during Covid-19, the first quarter of 2020 when the equity market was tremendously shocked. The reason that we purposely avoid using samples during that period is we want to alleviate the heteroskedasticity. An improvement could be made that identify the market's states, such as recession or booming, and make use of different Prior benchmark and different view strategy depending on the estimated market states.

3. Other optimisation methods could be applied, such as minimum VaR. One could also check the robustness by trying to implement constraints on optimisation.

4. One view is generated for each round of a loop. The LSTM model could be amended to generate several views. 
Also, the validity checking part could be included, and the confidence level could be treated as the uncertainty of views, as a further improvement of the model.

5. Different view generation methods could be used. The LSTM model could be replaced by any other algorithms or time-series models such as the GARCH.

6. Further robustness-checking processes could be performed by changing the date range, altering the prior, etc.

7. Tune a Model. There are several hyper-parameters that could be adjusted to improve the model's predictability and robustness. Since those hyperparameters are manually given and are not learned directly from the data during the training process, we may apply the grid search, random search, or a Bayesian optimization to tune Hyperparameter and improve the model performance. However, as our model is designed to train several models in the loop, that hyperparameter tunning process is expected to be extremely time-consuming, so I would not perform it here.

8. A more sophisticated model that can identify the current economic regimes could be devised. For example, we would like to design an algorithm that uses trading volume and volatility to classify market conditions. Then, we would use market conditions/states to confirm such as the prior. In the recession, we might use the Global Minimum Variance weights as the prior. In a booming market regime, a Maximum Sharpe Ratio weight might be applied. Similarly, the economic regimes could be used to identify such as the risk-aversion or hyper-parameter of views.

9. The optimisation process could include more variation, such as adding or releasing constraints, changing parameter $\lambda$ in the Markowitz mean-variance weights scheme, etc, for validity checking.

\section{Conclusion} \label{sec:Conclusion}

Overall, we present a portfolio construction process including the factors selection step and the weight allocation step. Three rationales are introduced as means of factors in our paper. 20 factors are selected, including equities and bonds from the US and Chinese markets, Commodities. Factors are proxied by characteristic ETFs. 

Then, various weight allocation schemes are provided to assign weights to each factor. A Deep Learning algorithm is applied to predict views. Then, we use the Black-Litterman model incorporating views with the Prior to get the Posterior.

In the end, the robustness checking shows how weights change with respect to time evolving and variance increasing. Results using shrinkage variance are also provided aiming to alleviate the impacts of representativeness of historical data, but there sadly has little impact. Overall, the model by using the Deep Learning plus Black-Litterman model results outperform the portfolio by other weight allocation schemes, even though further improvement and robustness checking should be performed.

\bibliographystyle{apalike}  
\bibliography{bibliography}

\begin{thebibliography}{}

\bibitem[Da~Silva et~al., 2009]{da2009black}
Da~Silva, A.~S., Lee, W., and Pornrojnangkool, B. (2009).
\newblock The black--litterman model for active portfolio management.
\newblock {\em The Journal of Portfolio Management}, 35(2):61--70.

\bibitem[Fama and French, 1993]{fama1993common}
Fama, E.~F. and French, K.~R. (1993).
\newblock Common risk factors in the returns on stocks and bonds.
\newblock {\em Journal of Financial Economics}, 33(1):3--56.

\bibitem[Fama and French, 2012]{fama2012size}
Fama, E.~F. and French, K.~R. (2012).
\newblock Size, value, and momentum in international stock returns.
\newblock {\em Journal of financial economics}, 105(3):457--472.

\bibitem[He and Litterman, 2002]{heLitterman2002intuition}
He, G. and Litterman, R. (2002).
\newblock The intuition behind black-litterman model portfolios.
\newblock {\em Available at SSRN 334304}.

\bibitem[Idzorek, 2007]{idzorek2007step}
Idzorek, T. (2007).
\newblock A step-by-step guide to the black-litterman model: Incorporating user-specified confidence levels.
\newblock In {\em Forecasting expected returns in the financial markets}, pages 17--38. Elsevier.

\bibitem[Kahn and Grinold, 1999]{kahn1999active}
Kahn, R. and Grinold, R. (1999).
\newblock Active portfolio management.
\newblock {\em A Quantitative Approach for Providing Superior Returns and Controlling Risk}.

\bibitem[Ledoit and Wolf, 2003]{ledoit2003improved}
Ledoit, O. and Wolf, M. (2003).
\newblock Improved estimation of the covariance matrix of stock returns with an application to portfolio selection.
\newblock {\em Journal of empirical finance}, 10(5):603--621.

\bibitem[Lin, 2020]{nasdaq}
Lin (2020).
\newblock A practitioner’s guide to multi-factor portfolio construction.
\newblock {\em Nasdaq Global Information Services}.

\bibitem[Maggiar, 2009]{maggiar2009active}
Maggiar, A. (2009).
\newblock Active fixed-income portfolio management using the black-litterman model.
\newblock {\em Available at SSRN 2655810}.

\bibitem[Satchell and Scowcroft, 2000]{RePEc:pal:assmgt:v:1:y:2000:i:2:d:10.1057_palgrave.jam.2240011}
Satchell, S. and Scowcroft, A. (2000).
\newblock A demystification of the black–litterman model: Managing quantitative and traditional portfolio construction.
\newblock {\em Journal of Asset Management}, 1(2):138--150.

\bibitem[Walters and Smith, 2014]{walters2014black}
Walters, J. and Smith, J. (2014).
\newblock The black-litterman model in detail.
\newblock {\em SSRN 1314585}, 10(2):100--120.

\end{thebibliography}

\end{document}